\documentclass[11pt,a4paper]{article}
\pdfoutput=1

\usepackage{jheppub}
\usepackage{rotating}

\usepackage{wrapfig}

\usepackage{amsfonts}
\usepackage{amsmath}
\usepackage{slashed}
\usepackage{amssymb}
\usepackage{latexsym}
\usepackage{multirow}
\usepackage{hyperref}
\usepackage{enumitem}
\hypersetup{colorlinks=true,citecolor=red,linkcolor=magenta,urlcolor=blue}
\usepackage[caption=false]{subfig}
\usepackage{relsize}
\usepackage{booktabs}
\usepackage{cleveref}
\usepackage{fancyhdr}
\usepackage{float}
\usepackage{graphicx}
\usepackage{microtype}
\usepackage{tabularx}
\usepackage{xcolor}

\makeatletter
\newcommand\footnoteref[1]{\protected@xdef\@thefnmark{\ref{#1}}\@footnotemark}
\makeatother

\newcolumntype{C}[1]{>{\centering\let\newline\\\arraybackslash\hspace{0pt}}m{#1}}

\newcommand{\LFU}{$R_{D^{(*)},K^{(*)}}$}

\definecolor{green}{rgb}{0.1,0.5,0.0}

\newcommand\inputpgf[2]{{
\let\pgfimageWithoutPath\pgfimage
\renewcommand{\pgfimage}[2][]{\pgfimageWithoutPath[##1]{#1/##2}}
\input{#1/#2}
}}

\title{The Flavourful Present and Future of 2HDMs at the Collider Energy Frontier}

\author[a]{Oliver Atkinson,}
\author[b]{Matthew Black,}
\author[a]{Christoph Englert,}
\author[b]{Alexander Lenz,}
\author[b]{Aleksey Rusov,}
\author[c]{James Wynne}

\affiliation[a]{SUPA, School of Physics \& Astronomy, University of Glasgow, Glasgow G12 8QQ, UK}

\affiliation[b]{Physik Department, Universit\"{a}t Siegen, Walter-Flex-Str. 3, 57068 Siegen, Germany}

\affiliation[c]{IPPP, Department of Physics, University of Durham, DH1 3LE, UK}

\emailAdd{o.atkinson.1@research.gla.ac.uk}
\emailAdd{Matthew.Black@uni-siegen.de}
\emailAdd{christoph.englert@glasgow.ac.uk}
\emailAdd{Alexander.Lenz@uni-siegen.de}
\emailAdd{rusov@physik.uni-siegen.de}
\emailAdd{jameswynne39@gmail.com}

\abstract{We study the intersection of flavour and collider physics for Two-Higgs-Doublet models of Type I and II. Drawing from the flavour precision-LHC exotics search complementarity, we also provide a projection of the future sensitivity that can be achieved in light of currently available analyses. On the one hand, we find that the parameter space of the 2HDM can be explored significantly further with more data from the LHC with some complementarity with flavour physics. On the other hand, flavour physics results alongside their projections remain powerful tools to constrain the model space in regions where direct sensitivity to new states via exotics searches is lost. Our results further highlight the recently observed flavour physics anomalies as important drivers of new physics searches in the future; we also touch on implications for a strong first order electroweak phase transition.}

\begin{document}
\hfill SI-HEP-2022-03

\maketitle

\section{Introduction}
\label{sec:introduction}
Searches for physics beyond the Standard Model (BSM) are key pillars of the current particle physics programme that stretches 
across many different experimental arenas. Shortcomings of the Standard Model (SM), e.g. the possible lack of enough 
CP violation or the insufficiently strong electroweak phase transition to address the Sakharov criteria~\cite{Sakharov:1967dj}, provide the motivation for extending the interaction and particle spectrum of the SM. Especially well-motivated scenarios along these lines are Higgs sector extensions. While the existence of a non-trivial electroweak vacuum structure had been verified with the discovery of the $W$ and $Z$ bosons at UA1/UA2, it was only in 2012 that the discovery of the Higgs boson~\cite{ATLAS:2012yve,CMS:2012qbp} confirmed the mechanism of spontaneous symmetry breaking~\cite{Englert:1964et, Higgs:1964ia, Higgs:1964pj} through the Higgs boson's couplings to the massive gauge bosons. Subsequent investigations of the 125~GeV Higgs boson after its discovery have shown that this new state closely follows the SM expectation. Nonetheless, current experimental constraints only limit modification of the Higgs boson's interactions to be below $\sim 10\%$~\cite{CMS:2018ipl,ATLAS:2020qdt} so there are reasonable margins for new, weakly-coupled modifications of the Higgs sector. Consistency with electroweak fits~\cite{Baak:2014ora} provides an additional constraint for extensions of the symmetry breaking Higgs sector that can be competitive with current direct measurements.

Out of the Higgs sector extensions that are typically considered for BSM investigations, two Higgs doublet models (2HDMs) are particularly well-motivated theories (for reviews see~Refs.~\cite{Branco:2011iw,Gunion:1989we}). On the one hand, they implement custodial isospin analogous to the SM Higgs field, thus avoiding tensions and fine-tuning with electroweak precision measurements that occur in higher dimensional representations of electroweak $SU(2)_L\times U(1)_Y$ breaking~\cite{Gunion:1990dt}. On the other hand, 2HDMs can be considered as harbingers of supersymmetric theories, which demand a second Higgs doublet due to holomorphy, and (non-)perturbative anomaly cancellation to maximise the $S$-matrix symmetry. 

Leaving aside the theoretical motivation for introducing a second weak doublet contributing to electroweak symmetry breaking and fermion mass generation, the experimental avenues to constrain or even observe a 2HDM extension of the SM are plentiful: links of fermion vs. gauge boson mass generation provide a motivation to combine precision flavour physics investigations with those performed at the high energy frontier, chiefly at the Large Hadron Collider (LHC). In fact, a recent investigation into the flavour and Higgs signal strength constraints of the 2HDM~\cite{Atkinson:2021eox} demonstrated the power of flavour measurements in driving BSM mass scales into a region where the direct detection at the LHC could become challenging.
Additionally, current flavour anomalies highlight flavour physics as a particularly relevant area to inform future investigations at the high energy frontier.

Building on~Ref.~\cite{Atkinson:2021eox}, we investigate the intersection of flavour and collider physics for the 2HDMs of type I and II in this work, with a view towards the High Luminosity (HL) LHC phase. Clarifying and extrapolating the currently available search strategies in flavour physics and for the exotic states that are predicted in the 2HDM enables us to discuss flavour-collider complementarity and identify regions that require targeting to enhance the BSM potential during the upcoming LHC runs.

We organise this work as follows: in Section~\ref{sec:model}, we review the basics of the 2HDM to make this work self-consistent. Section~\ref{sec:theory} provides details on theoretical and electroweak constraints that set the baseline of our investigation, then in Section~\ref{sec:SS} we discuss the status of the SM Higgs signal strengths in the 2HDM type~I (2HDM-I). In Section~\ref{sec:flavour} we discuss in detail the flavour constraints on the 2HDM-I, extending the work of~Ref.~\cite{Atkinson:2021eox}, before we move on to contrast LHC searches and measurements in~Section~\ref{sec:lhc} for the 2HDM-I and 2HDM-II. We also comment on expected improvements of both flavour and collider constraints before discussing cosmological implications in Section~\ref{sec:cosmo}. We provide a summary and conclusions in Section~\ref{sec:conclusion}.

\section{The Two Higgs Doublet Model}
\label{sec:model}
We follow the notation of Refs.~\cite{Branco:2011iw,Gunion:1989we}, where, instead of using the SM Higgs doublet and its (symplectic) charge-conjugated version to gain mass terms for both up-type and down-type quarks, one can introduce two distinct doublets
\begin{equation}
    \Phi_i = \begin{pmatrix} \phi_i^+ \\ (v_i+\phi_i^0 + iG_i^0)/\sqrt{2} \end{pmatrix},
\end{equation}
with $i=1,2$. 
After electroweak symmetry breaking (EWSB), these doublets lead to 5 physical Higgs bosons (the other 3 degrees of freedom construct the longitudinal $W^\pm$, $Z^0$ boson polarisations).
One finds two charged Higgs, $H^\pm$, two neutral scalars, $h^0,H^0$, and a neutral pseudoscalar, $A^0$. 
We take $h^0$ to be the lighter of the two neutral scalars and the currently observed Higgs particle with mass $125.25\pm0.17\,$GeV \cite{ParticleDataGroup:2020ssz}.
The potential for a general 2HDM with a softly broken $\mathbb{Z}_2$ symmetry is \cite{Branco:2011iw}
\begin{equation}
\begin{split}
    V(\Phi_1,\Phi_2)=m_{11}^2\Phi_1^\dagger\Phi_1+m_{22}^2\Phi_2^\dagger\Phi_2
    -m_{12}^2(\Phi_1^\dagger\Phi_2+\Phi_2^\dagger\Phi_1)
    + \frac{\lambda_1}{2}(\Phi_1^\dagger\Phi_1)^2+\frac{\lambda_2}{2}(\Phi_2^\dagger\Phi_2)^2 \\ +
    \lambda_3 (\Phi^ \dagger_1\Phi_1) (\Phi^\dagger_2\Phi_2) + \lambda_4 (\Phi^\dagger_1\Phi_2) (\Phi^\dagger_2\Phi_1) +
    \frac{\lambda_5}{2} \left[(\Phi^\dagger_1\Phi_2)^2+(\Phi^\dagger_2\Phi_1)^2\right].
\end{split}
\end{equation}
We can use this potential in its ``lambda basis'' to construct mass terms for each of the physical Higgs particles, which, alongside the vacuum expectation value (VEV) $v^2=v_1^2+v_2^2$, the ratio of VEVs $\tan\beta=v_2/v_1$, the mixing parameter $\cos(\beta-\alpha)$, and the softly $\mathbb{Z}_2$ breaking term $m_{12}$ form the ``mass basis'' which more simply translates to physical observables, see e.g.~Refs.~\cite{Atkinson:2021eox,Basler:2016obg,Arnan:2017lxi,Han:2020zqg,Kling:2016opi} for transformations between bases.

The Yukawa couplings in the 2HDM can be generally expressed as
\begin{equation}
{\cal L}_Y = 
-y_{ij}^1 \bar{\Psi}^i_L \Phi_1 \psi^j_R
- 
y_{ij}^2 \bar{\Psi}^i_L \Phi_2 \psi^j_R \, .
\end{equation}
In general the 2HDM Yukawa couplings need not be flavour diagonal and so it is possible to allow tree-level flavour-changing neutral currents (FCNCs) which are absent in 
the~SM. 
In order to maintain this natural flavour conservation, one can consider fermions coupling to the Higgs doublets in specific ways. 
The four types of 2HDM with natural flavour conservation are described in Table~\ref{tab:types}.
\begin{table}[!t]
    \centering
    \begin{tabular}{|c||c|c|c|c|}
        \hline
        \rm Model & I & II & X & Y \\
        \hline\hline
        $u^i_R$ & $\Phi_2$  & $\Phi_2$  & $\Phi_2$  & $\Phi_2$ \\
        \hline
        $d^i_R$ & $\Phi_2$  & $\Phi_1$  & $\Phi_2$  & $\Phi_1$ \\
        \hline
        $e^i_R$ & $\Phi_2$  & $\Phi_1$  & $\Phi_1$  & $\Phi_2$ \\
        \hline
    \end{tabular}
    \caption{Types of 2HDM with natural flavour conservation from demanding fermions couple to specific doublets.} 
    \label{tab:types}
\end{table}
After spontaneous symmetry breaking, we can write the Yukawa sector of the 2HDM Lagrangian as
(see e.g. Refs.~\cite{Crivellin:2019dun, Branco:2011iw}) 
\begin{eqnarray}
    \nonumber
    {\cal L}^{\rm 2HDM}_{\rm Yukawa} & = & 
    - \sum \limits_{f= u,d,\ell} \frac{m_f}{v} 
    \left(\xi_h^f \, \bar{f} f h + \xi_H^ f \, \bar{f} f H - i \xi_A^f  \, \bar{f} \gamma_5 f A \right) \\
    & - & \left[
        \frac{\sqrt{2} V_{ud}}{v} \bar{u} 
        \left(m_d \, \xi_A^d P_R - m_u \, \xi_A^u P_L \right) d H^+ 
        + \frac{\sqrt{2}}{v} m_\ell \, \xi_A^l  (\bar{\nu}  P_R  \ell)  H^+ + {\rm h.c.}
        \right] \! .
        \label{eq:yukawa}
\end{eqnarray}
Imposing the fermion interactions to specific doublets in Table~\ref{tab:types}, the $\xi$ coupling strengths are expressed in Table~\ref{tab:xis}.\footnote{Note that we differ from Ref.~\cite{Branco:2011iw} by a sign in the $H^+$ Yukawa interaction which in turn leads to a difference in the fermion coupling strengths. This is the same convention as found in e.g. Ref.~\cite{Crivellin:2019dun}.}
\begin{table}[th]
    \centering
    \begin{tabular}{|c||c|c|c|c|}
        \hline
        \rm Model & I & II & X & Y \\
        \hline\hline
        $\xi^u_h$ & $\cos\alpha/\sin\beta$ & $\cos\alpha/\sin\beta$ & $\cos\alpha/\sin\beta$ & $\cos\alpha/\sin\beta$ \\
        \hline
        $\xi^d_h$ & $\cos\alpha/\sin\beta$ & $-\sin\alpha/\cos\beta$ & $\cos\alpha/\sin\beta$ & $-\sin\alpha/\cos\beta$ \\
        \hline
        $\xi^l_h$ & $\cos\alpha/\sin\beta$ & $-\sin\alpha/\cos\beta$ & $-\sin\alpha/\cos\beta$ & $\cos\alpha/\sin\beta$ \\
        \hline\hline
        $\xi^u_H$ & $\sin\alpha/\sin\beta$ & $\sin\alpha/\sin\beta$ & $\sin\alpha/\sin\beta$ & $\sin\alpha/\sin\beta$ \\
        \hline
        $\xi^d_H$ & $\sin\alpha/\sin\beta$ & $\cos\alpha/\cos\beta$ & $\sin\alpha/\sin\beta$ & $\cos\alpha/\cos\beta$ \\
        \hline
        $\xi^l_H$ & $\sin\alpha/\sin\beta$ & $\cos\alpha/\cos\beta$ & $\cos\alpha/\cos\beta$ & $\sin\alpha/\sin\beta$ \\
        \hline\hline
        $\xi^u_A$ & $\cot\beta$ & $\phantom{-}\cot\beta$ & $\phantom{-}\cot\beta$ & $\phantom{-}\cot\beta$ \\
        \hline
        $\xi^d_A$ & $\cot\beta$ & $-\tan\beta$ & $\phantom{-}\cot\beta$ & $-\tan\beta$ \\
        \hline
        $\xi^l_A$ & $\cot\beta$ & $-\tan\beta$ & $-\tan\beta$ & $\phantom{-}\cot\beta$ \\
        \hline
    \end{tabular}
    \caption{Coupling strengths $\xi$ in each type of 2HDM between the Higgs particles and fermions.}
    \label{tab:xis}
\end{table}

\section{Theory and Electroweak Precision Constraints}
\label{sec:theory}
\subsection{Perturbativity, Unitarity, and Vacuum Stability}
Perturbativity in the scalar sector can be simply expressed as \cite{Chen:2018shg,Ginzburg:2005dt} 
\begin{equation}
    |\lambda_i|\leq 4\pi, \qquad 
    i = 1, \ldots 5.
    \label{perturbativity}
\end{equation}
As in Ref.~\cite{Atkinson:2021eox} we also consider the less conservative bounds of 
$ |\lambda_i|\leq 4$ -- inspired by the results in 
Refs.~\cite{Grinstein:2015rtl,Cacchio:2016qyh}.
Looking at the couplings in Eq.~(\ref{eq:yukawa}), one can further consider 
the perturbativity constraints from the Yukawa sector:
\begin{equation}
\begin{aligned}
\frac{\sqrt{2} \, V_{tb} \, m_t \cot \beta}{2 v} \le { \sqrt{4 \pi}} & \quad \Longrightarrow \quad \tan \beta > {0.14} \, ,
\label{eq:pert_Yukawa}
\end{aligned}
\end{equation}
where we have chosen a very conservative range.
For the range of $\log [\tan \beta]$ we will consider the conservative lower bound 
$\tan \beta = 10^{-1.5} \approx 0.03$. For the upper bound 
we adopt the same $\tan \beta = 10^{+2.5} \approx 300$ 
as in Ref.~\cite{Atkinson:2021eox}, for ease of comparison of results 
between 2HDM-II~\cite{Atkinson:2021eox} and 2HDM-I (this work).

We apply the conditions for a stable vacuum as set out in Ref.~\cite{Deshpande:1977rw}, whilst demanding the vacuum to be the global minimum of the potential \cite{Barroso:2013awa}. We also consider the conditions from tree-level unitarity, see Refs.~\cite{Ginzburg:2005dt,Arhrib:2000is}\footnote{{ For a similar discussion in an alternative lambda basis see Ref.~\cite{Horejsi:2005da}.} },
alongside NLO unitarity and the condition that NLO corrections to partial wave amplitudes are suppressed relative to LO contributions, see  Refs.~\cite{Cacchio:2016qyh,Grinstein:2015rtl}. 

We get the same constraints on the 2HDM parameters as found in our previous study from theory, where these are not dependent on the specific couplings defining the different types of 2HDM. 
We refer to Fig.~1 of Ref.~\cite{Atkinson:2021eox} for these results, where the main implication is a close mass degeneracy for the new Higgses enforced from $\sim1\,$TeV and becomes stricter as the mass scale increases.

\subsection{Oblique corrections}
The expressions for the electroweak precision observables (EWPOs): $S$, $T$, and $U$~\cite{Peskin:1990zt,Peskin:1991sw} (also referred to as oblique corrections) in the 2HDM (derived from Ref.~\cite{Grimus:2008nb}) are explicitly given in Appendix~C of
Ref.~\cite{Atkinson:2021eox}. New physics generally has only a small effect on $U$ in comparison to the $S$ and $T$
parameters as the latter correspond to distinct dimension 6 operators in 
the effective field theory expansion, in contrast to $U$~\cite{Barbieri:2004qk}. 
It is therefore justified to follow the approach that neglects $U$ from fits 
by setting $U=0$ as outlined in Ref.~\cite{Baak:2014ora,ParticleDataGroup:2020ssz}. 
This effectively reduces the error on the experimental result for $T$, 
due to the correlation between $T$ and $U$.
The oblique parameters will be included in our global fit.

\section{Higgs Signal Strengths}
\label{sec:SS}
Many measurements of the properties of the observed Higgs boson have been made. Of these, the most relevant for our discussion are the Higgs signal strengths $\mu^f_i$, which are defined as a ratio between experimental and SM values of the product of the cross section and branching fraction in a given channel with production mode $i$ and decay products $f$:
\begin{equation}
    \mu^f_i = \frac{(\sigma_i \cdotp {\cal B}_f)_\text{Exp.}}{(\sigma_i \cdotp {\cal B}_f)_\text{SM}}.
\end{equation}
Measurements of $\mu^f_i$ essentially serve as an indicator of how closely the observed Higgs matches 
the SM expectations. We include results from 31 channels 
\cite{CMS:2018uag, CMS:2019hve, ATLAS:2019nkf, ATLAS:2020qcv, ATLAS:2020fzp, CMS:2020gsy, ATLAS:2020qdt}
(collected in Table~4 of Ref.~\cite{Atkinson:2021eox}).

\begin{figure}[th]
	\centering
	\includegraphics[width=0.48\textwidth]{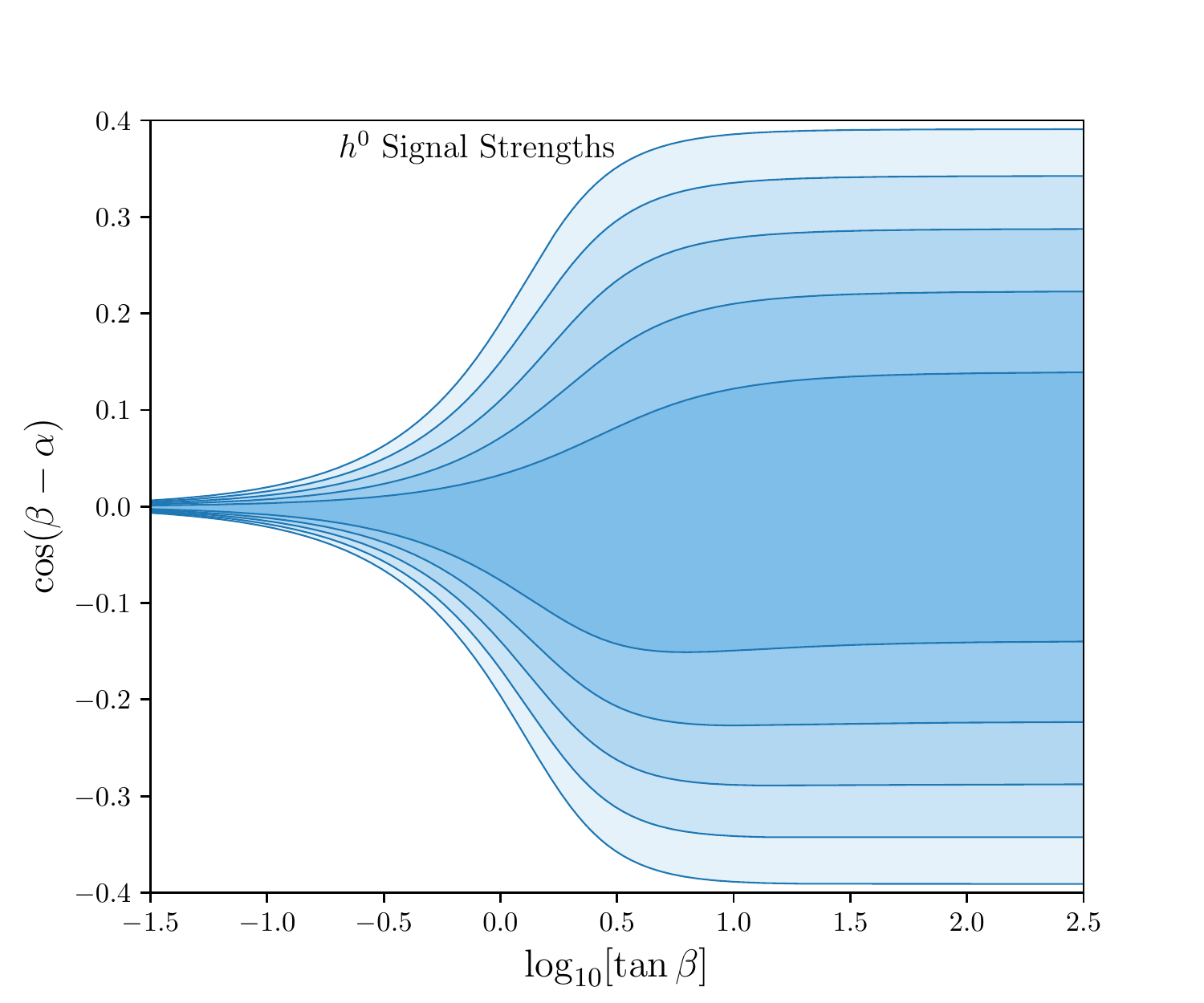}
	\caption{Combined fits of the SM Higgs signal strengths in the 2HDM-I (in the ($\tan\beta-\cos(\beta-\alpha)$) plane). Contours are shown representing allowed parameter space at 1, 2, 3, 4, 5$\sigma$ confidence from darkest to lightest.}
	\label{fig:SS}
\end{figure}

If the observed Higgs is not that of the SM, but part of an extended Higgs sector such as the 2HDM, its couplings to other particles will differ from the SM, a difference that will feed through into the values of the signal strengths. This is the case in the 2HDM-I, with the couplings being modified by the factors shown in Table~\ref{tab:xis}. It is therefore possible to use the signal strength measurements to constrain the parameters on which these couplings depend, namely the mixing angles. We do this by performing a fit using analytical calculations of the signal strengths in the 2HDM-I as functions of the input parameters $\tan\beta$ and $\cos(\beta-\alpha)$ from expressions given in Ref.~\cite{Gunion:1989we}, with results presented in Fig.~\ref{fig:SS}. 
Opposed to the 2HDM-II case (see Ref.~\cite{Atkinson:2021eox}) 
we find that more sizeable deviations from the alignment limit $\cos (\beta - \alpha) = 0$
are allowed for the 2HDM-I. We observe that larger deviations from the alignment limit 
are possible with increasing $\tan\beta$, and we find $|\cos (\beta - \alpha)| \leq 0.21\;(0.4) $ 
at $2\;(5)\sigma$ for sufficiently large $\tan \beta \geq 10$.

\section{Flavour Constraints in 2HDM-I}
\label{sec:flavour}
In Section~\ref{sec:flavour_fits}, we cover the constraints on the 2HDM-I parameter space from flavour observables, and then in Section~\ref{sec:global_fit} we combine these with the Higgs signal strengths and the EWPOs.
We follow the procedure of the earlier study \cite{Atkinson:2021eox} in the 2HDM-II, where we now exchange the Type II couplings for the Type I. 
All observables we consider here are listed with SM predictions and experimental measurements in Tables 4-8 of Ref.~\cite{Atkinson:2021eox}; the measurements are collected from various results throughout literature 
\cite{ATLAS:2020qdt,CMS:2020gsy,CMS:2021kom,ATLAS:2021upe,ATLAS:2020qcv,ATLAS:2020fzp,CMS:2020xwi,ATLAS:2020bhl,CMS:2018uag,ATLAS:2020fcp,CMS:2019hve,ParticleDataGroup:2020ssz,HFLAV:2019otj,LHCb:2020cyw,LHCb:2021qbv,McLean:2019qcx,Altmannshofer:2021qrr,LHCb:2014cxe,LHCb:2020gog,LHCb:2020lmf,ATLAS:2018gqc,CMS:2015bcy,CMS:2017rzx,BELLE:2019xld,Belle:2019oag,LHCb:2015wdu,BaBar:2013qry,Belle:2016fev,LHCb:2015ycz,CMS:2018qih,LHCb:2015tgy,LHCb:2018jna,LHCb:2021trn,LHCb:2014vgu,LHCb:2017avl,BaBar:2012mrf}
and the SM predictions are calculated in {\bf flavio} unless otherwise stated.

\subsection{Individual Fits}
\label{sec:flavour_fits}
For the tree-level (semi-)leptonic flavour-changing charged transitions, the 2HDM-I contributes through the effective operators
\begin{align}
    \label{eq:O-SP}
    \mathcal{O}_{S - P} = (\bar{u} P_L \, d) (\bar{\ell} P_L \nu_{\ell}), \qquad
    \mathcal{O}_{S + P} = (\bar{u} P_R \, d) (\bar{\ell} P_L \nu_{\ell}),
\end{align} 
with $P_{L(R)} = (1 \mp \gamma_5)/2$.
The Wilson coefficients of these operators in terms of the 2HDM-I parameters are
\begin{equation}
    C_{S-P} = -\frac{m_u \, m_\ell \cot^2\beta}{m_{H^+}^2}, \qquad\quad
    C_{S+P} = \frac{m_d \, m_\ell \cot^2\beta}{m_{H^+}^2} \, .
    \label{eq:dulnu}
\end{equation}
For a description of all operators gaining contributions from the 2HDM for the observables we consider and the translation of basis for these, we refer to \cite{Atkinson:2021eox}, taking expressions for the Wilson coefficients from Refs.~\cite{Crivellin:2019dun,Ilisie:2015tra,Borzumati:1998nx} where these are expressed in terms of general 2HDMs and one can simply insert the appropriate couplings for the 2HDM-I.

\begin{figure}[t]
	\centering
	\includegraphics[width=0.48\textwidth]{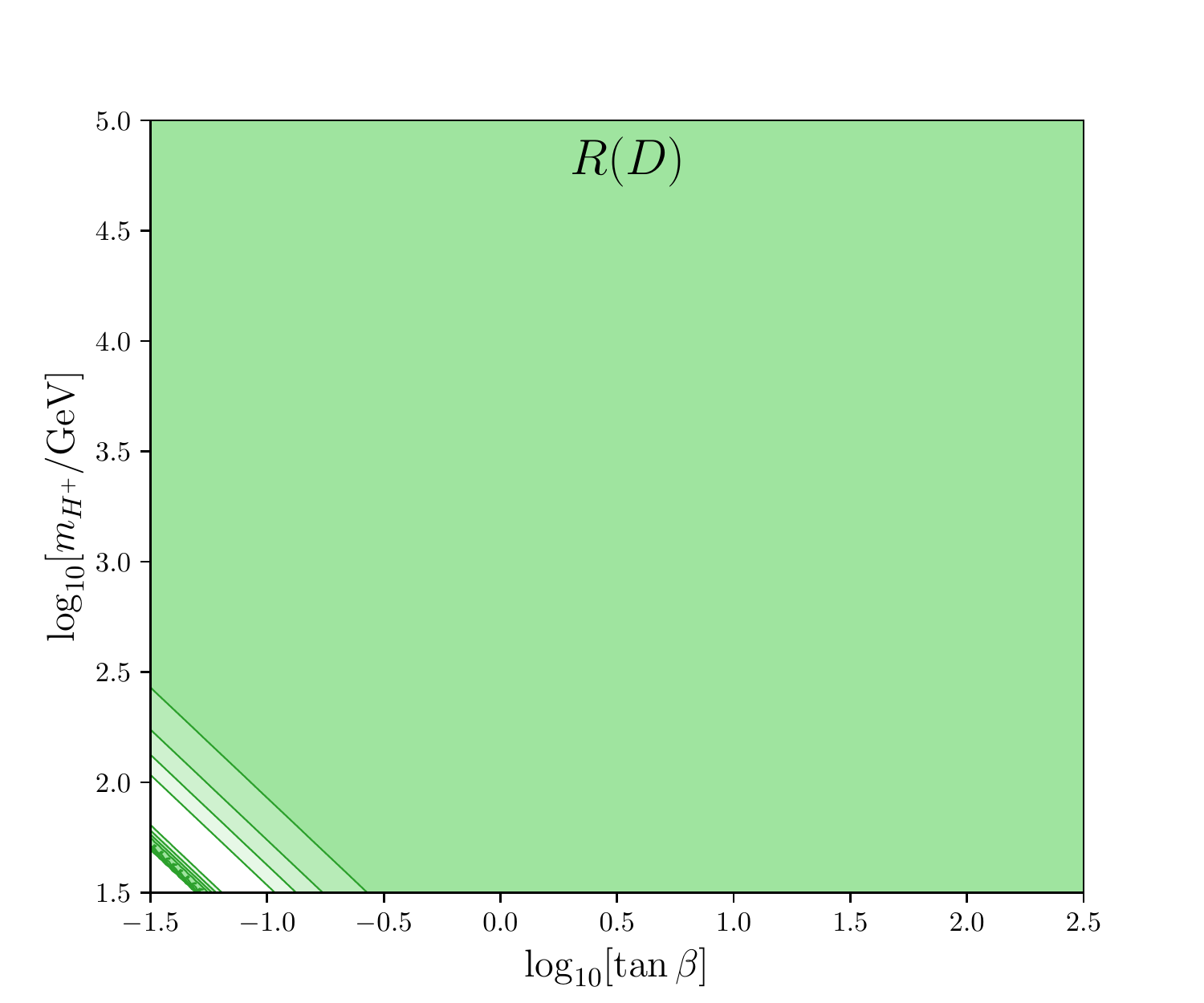}
	\includegraphics[width=0.48\textwidth]{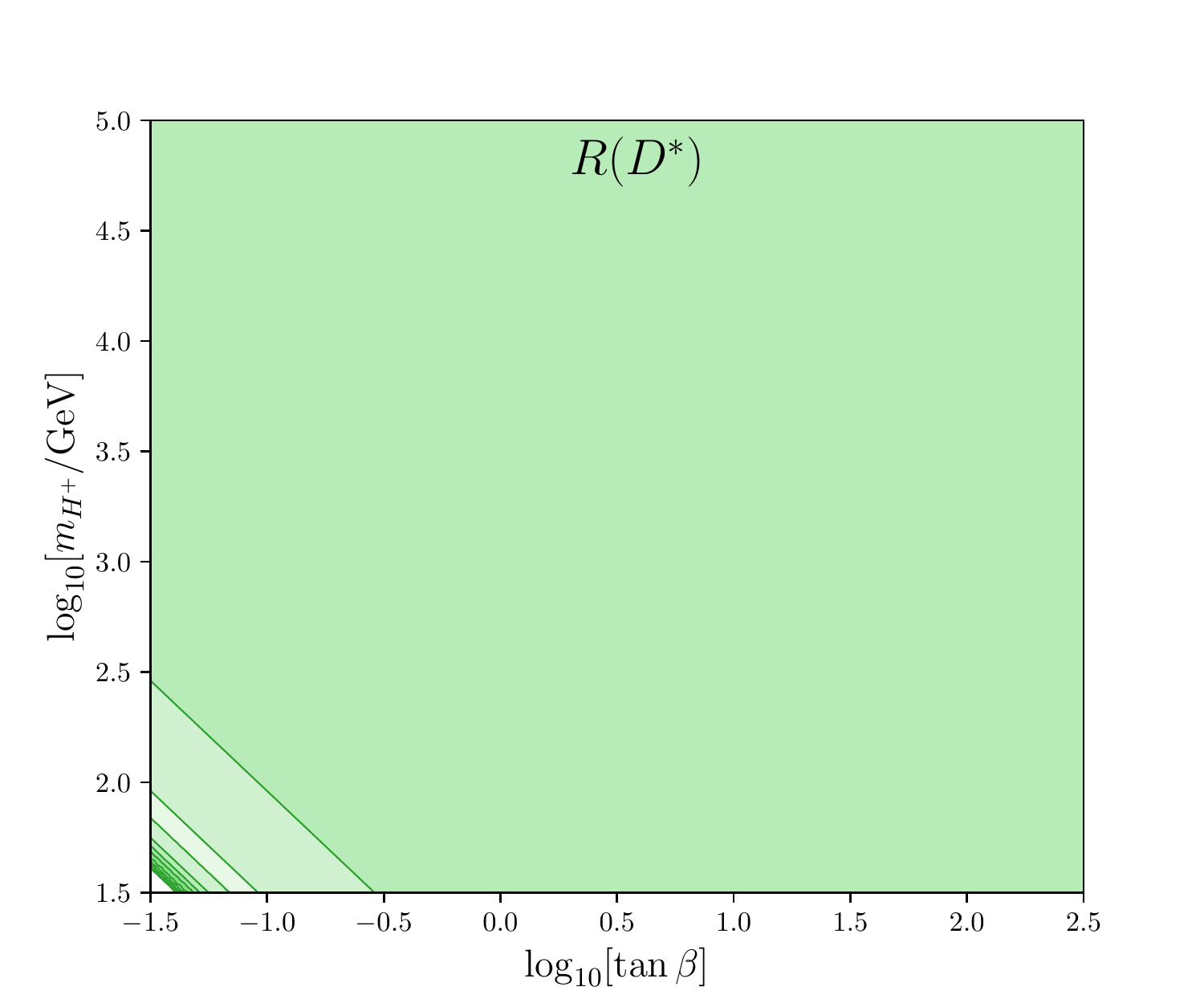} 
	\caption{Contour plots of the allowed 2HDM-I parameter space in the ($\tan\beta-m_{H^+}$) plane for $R(D)$ (left) and $R(D^*)$ (right). Contours are shown representing allowed parameter space at 1, 2, 3, 4, 5$\sigma$ confidence from darkest to lightest.}
	\label{fig:RD}
\end{figure}

\begin{figure}[t]
	\centering
	\includegraphics[width=0.48\textwidth]{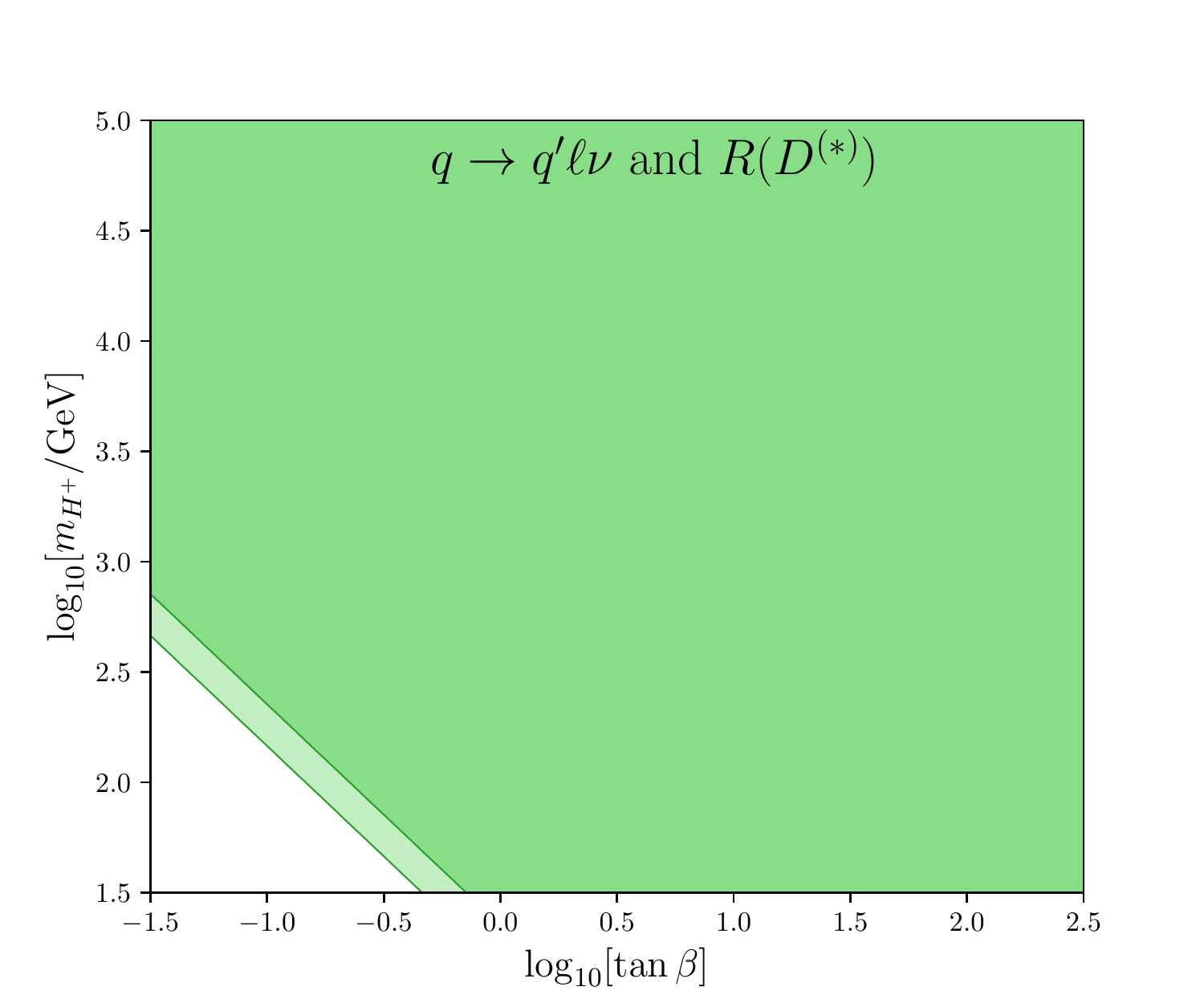}
	\caption{Contour plot of the allowed 2HDM-I parameter space in the ($\tan\beta-m_{H^+}$) plane for the combination of tree-level leptonic and semi-leptonic decays of $B, B_s, D, D_s, K$, and $\pi$ mesons and the hadronic decays of $\tau$ leptons to $K$ and $\pi$ mesons with a tau neutrino as well as $R(D)$ and $R(D^*)$. The lighter contour indicates the allowed parameter space at $2\sigma$ confidence level while the darker contour corresponds to $1\sigma$.}
	\label{fig:L-and-SL}
\end{figure}

First we consider the Lepton-Flavour Universality (LFU) observables 
$R(D^{(*)})\equiv$ \\${\cal B} (B \to D^{(*)} \tau \bar \nu_\tau)/{\cal B} (B \to D^{(*)} \ell \bar \nu_\ell)$, 
where $\ell = e,\mu$.
The latter implies a $2.8\sigma$ tension between experimental measurements and SM predictions, and the combined tension of the two between experiment and the SM is $3.2\sigma$ (using {\bf flavio}).
We show in Fig.~\ref{fig:RD} the allowed 2HDM-I parameter space for these two observables in individual fits. 
Within the theoretically-motivated parameter limits of our fits, we find that $R(D)$ allows a large part of our parameter space within $2\sigma$, whereas $R(D^*)$ allows most of our parameter space at $2.8\sigma$ or above.
In fact, for both observables, we find the best fit for $m_{H^+}\sim 1\,$GeV which is outside the physical domain of our model.
Within the considered limits of our parameters, we find a minimum of $3.5\sigma$ combined tension between the 2HDM-I and experiment for $R(D^{(*)})$.
In~Fig.~\ref{fig:L-and-SL}, we present the combined fit of tree-level flavour-changing charged currents including leptonic and semi-leptonic decays of mesons, hadronic decays of $\tau$ leptons, and $R(D^{(*)})$; see Table 6 of Ref.~\cite{Atkinson:2021eox} for all channels considered. 
The SM predictions in {\bf flavio} for the leptonic and semi-leptonic channels are based on Refs.~\cite{Bernlochner:2017jka,Caprini:1997mu,Sakaki:2013bfa,Bharucha:2015bzk,Gubernari:2018wyi,Detmold:2016pkz,Bernard:2006gy,Bernard:2009zm,FlaviaNetWorkingGrouponKaonDecays:2010lot}, and the new 2HDM-I contributions are described by Eq.~\eqref{eq:dulnu}.

\begin{figure}[t]
    \centering
    \includegraphics[width=0.48\textwidth]{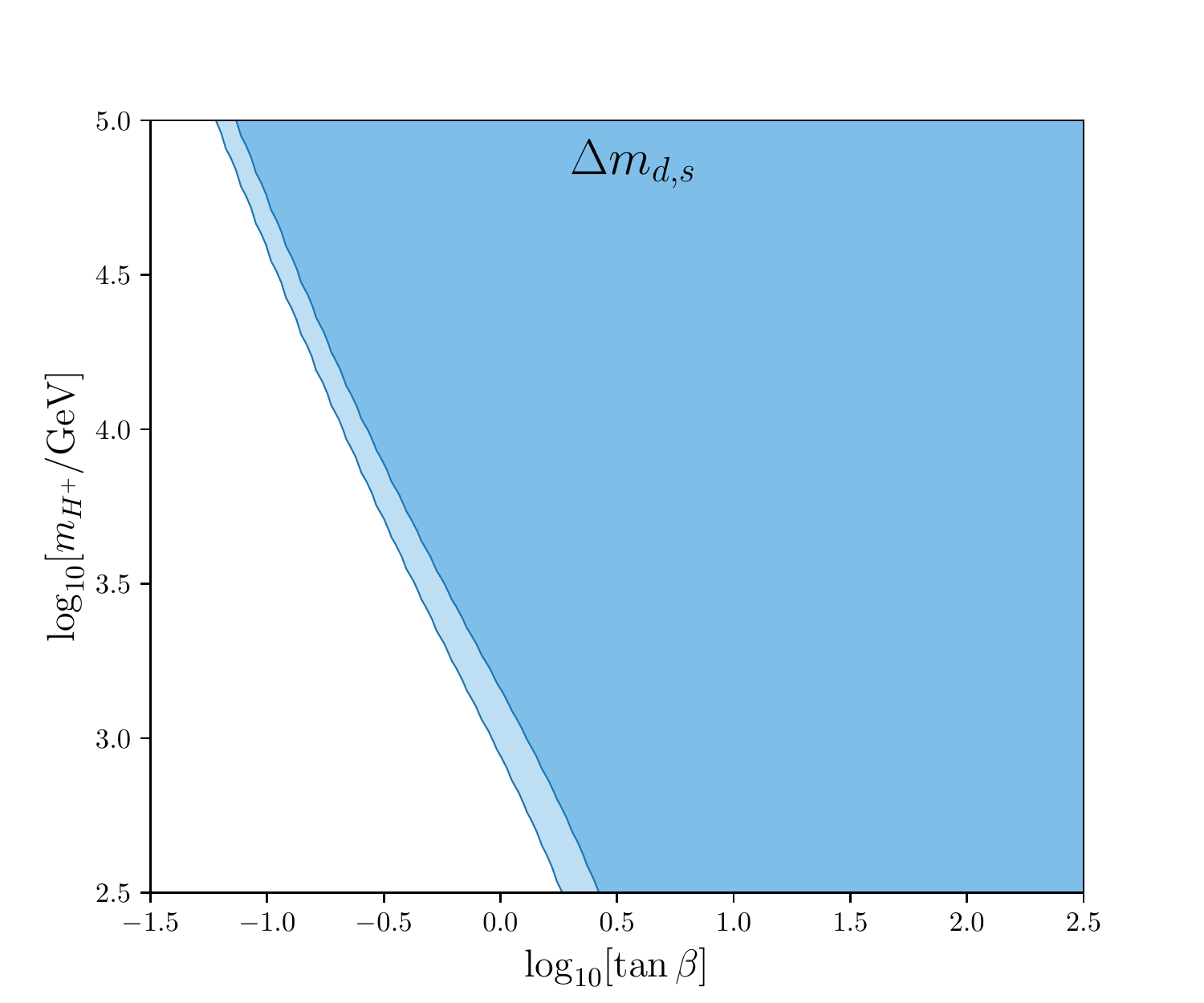}
    \caption{Contour plot of the allowed 2HDM-I parameter space in the ($\tan\beta-m_{H^+}$) plane for the combined fit to $\Delta m_{d,s}$. The darker contour indicates allowed parameter space at $1\sigma$ confidence, and the lighter at $2\sigma$.}
    \label{fig:mixing}
\end{figure}

Next we consider the mixing of neutral $B_{d,s}$ mesons.
The mass differences of this meson mixing are experimentally known very precisely at the level~$\sim{\cal O}(0.1\%)$ \cite{ParticleDataGroup:2020ssz}. On the theory side, however, the uncertainties are still dominated by the non-perturbative determinations of the matrix elements of the $\Delta B=2$ operators.
We use the averages presented in Ref.~\cite{DiLuzio:2019jyq} combining HQET Sum Rules \cite{King:2019lal,Kirk:2017juj,Grozin:2016uqy} and lattice calculations \cite{Dowdall:2019bea,Boyle:2018knm,FermilabLattice:2016ipl}, yielding a theory precision of~${\cal O}(5\%)$.
Terms proportional to the $tH^-$ coupling dominate the 2HDM contributions in the box diagrams with the down-type quark effects largely mass-suppressed. 
The predictions in the 2HDM-I for $B$-meson mixing are therefore very similar to those in the 2HDM-II where the dominating up-type $H^\pm$ couplings remain the same. 
The combined fit for $\Delta m_{d,s}$ is shown in Fig.~\ref{fig:mixing}, where we find a correlation between $\tan\beta$ and $m_{H^+}$ constraining $\tan\beta$ from below.
We take the expressions for the $\Delta B=2$ Wilson coefficients from 
Ref.~\cite{Crivellin:2019dun}, converting to the {\bf flavio} basis as described in Ref.~\cite{Atkinson:2021eox}.

\begin{figure}[th]
    \centering
    \includegraphics[width=0.48\textwidth]{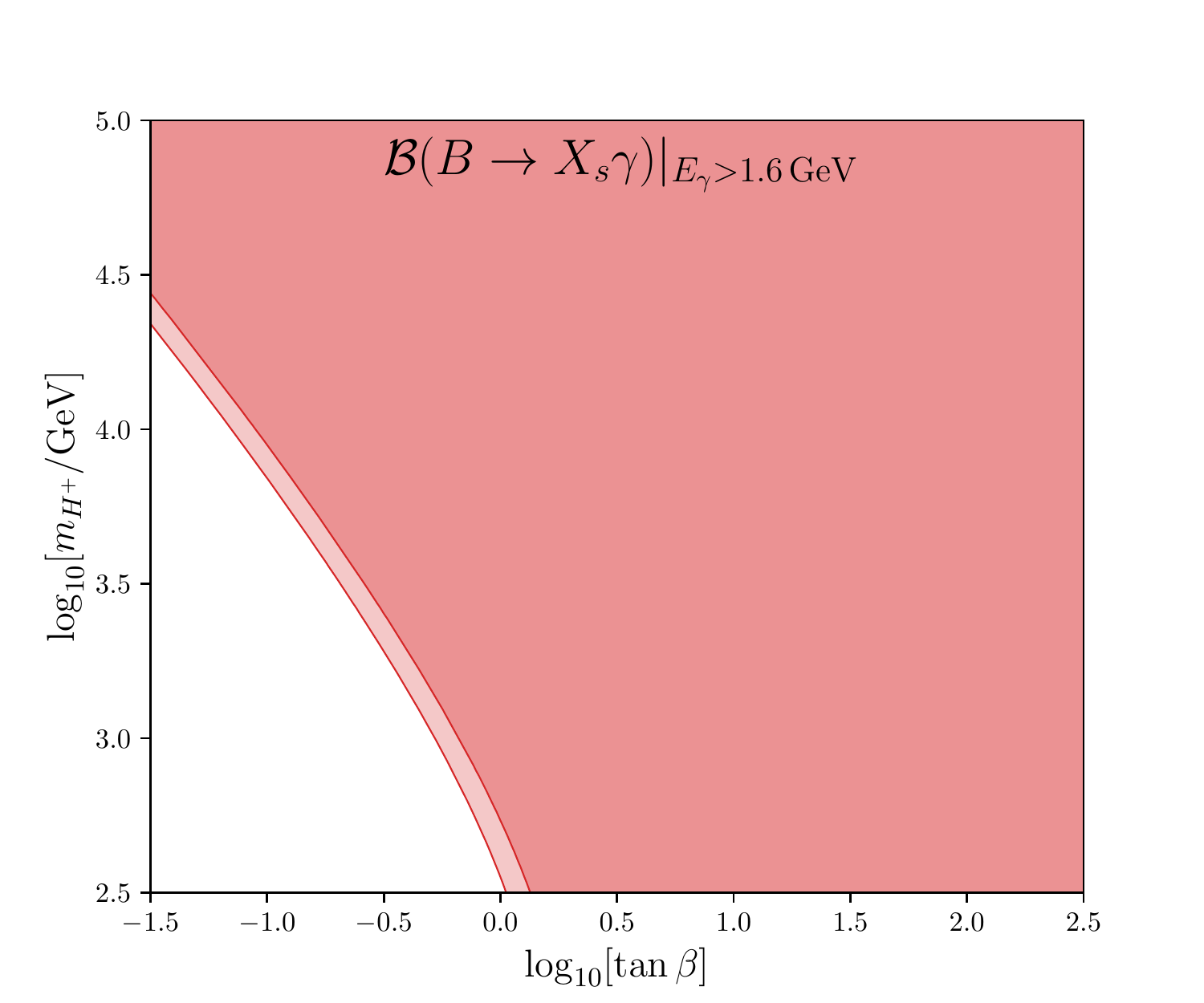}
    \caption{Contour plot of the allowed 2HDM-I parameter space in the ($\tan\beta-m_{H^+}$) plane for ${\cal B}(\bar B\to X_s\gamma)$. The darker contour indicates allowed parameter space at $1\sigma$ confidence, and the lighter at $2\sigma$.}
    \label{fig:bsgamma}
\end{figure}

A signature observable for new physics analysis is the branching ratio of the $\bar B\to X_s\gamma$ decay. 
In the SM, this is known to NNLO in QCD \cite{Misiak:2020vlo} (based on the previous Refs.~\cite{Misiak:2006ab,Misiak:2015xwa}).
We take the experimental average \cite{HFLAV:2019otj} formed from Refs.~\cite{CLEO:2001gsa,BaBar:2012fqh,Belle:2016ufb}.
In the 2HDM-II, $\bar{B}\to X_s\gamma$ provides a distinct constraint on the lower bound for the charged Higgs mass $m_{H^+}$; in the Type~I however, this is much more correlated with $\tan\beta$, and we therefore cannot find a clear constraint from this observable alone.
In Fig.~\ref{fig:bsgamma} we show the fit to ${\cal B}(\bar{B}\to X_s\gamma)|_{E_\gamma>1.6\,{\rm GeV}}$ in the 2HDM-I where we calculate the contributions to the Wilson coefficents $C_7,C_8$ at NLO using Ref.~\cite{Borzumati:1998nx}.

\begin{figure}[!t]
    \centering
    \includegraphics[width=0.48\textwidth]{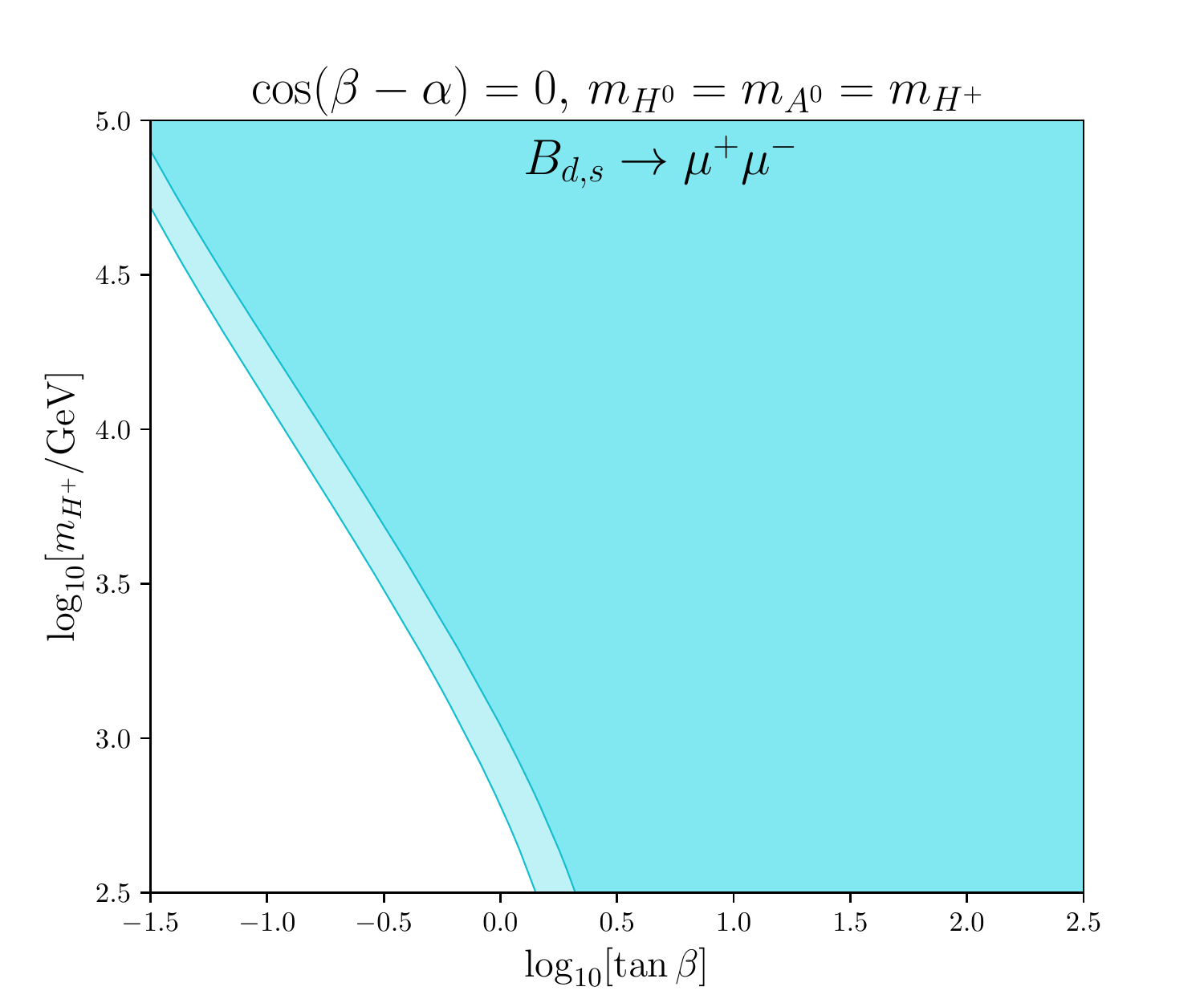}
    \caption{Contour plot of the allowed 2HDM-I parameter space in the ($\tan\beta-m_{H^+}$) plane for the combined fit to ${\cal}(B_{d,s}\to\mu^+\mu^-)$, taken in the limits of alignment ($\cos(\beta-\alpha)=0$) and degenerate masses ($m_{H^0}=m_{A^0}=m_{H^+}$). The darker contour indicates allowed parameter space at $1\sigma$ confidence, and the lighter at $2\sigma$.}
    \label{fig:bsmumu}
\end{figure}

We also consider the FCNC processes $B_{d,s}\to\mu^+\mu^-$, which are sensitive to BSM contributions to scalar operators, making them important observables to test the 2HDM. 
In recent years, ATLAS, CMS, and LHCb \cite{LHCb:2017rmj,ATLAS:2018cur,CMS:2019bbr,LHCb:2020zud,LHCb:2021awg,LHCb:2021vsc} have all produced measurements for $B_s\to\mu^+\mu^-$, and developed further the upper limit on $B_d\to\mu^+\mu^-$.
In our analysis we make use of the combination of these measurements from Ref.~\cite{Altmannshofer:2021qrr}.
The theory prediction is formed of the perturbative calculations \cite{Buchalla:1993bv,Bobeth:2013uxa,Beneke:2019slt} and the determinations of the non-perturbative decay constants, for example \cite{ETM:2016nbo,Bazavov:2017lyh,Hughes:2017spc}.
The 2HDM Wilson coefficients (taken again from Ref.~\cite{Crivellin:2019dun}) for the operators contributing to these processes, ${\cal O}_{10}^{(')},{\cal O}_{S}^{(')},{\cal O}_{P}^{(')}$, also depend on further 2HDM parameters:
$\cos(\beta - \alpha), m_{H^0}, m_{A^0}$. 
Motivated by constraints from theory and the Higgs signal strengths 
(see Sections \ref{sec:theory} and \ref{sec:SS}), we present in Fig.~\ref{fig:bsmumu} 
a 2D fit in the ($\tan\beta-m_{H^+}$)-plane fixing $\cos(\beta-\alpha)=0$ 
and $m_{H^0} = m_{A^0} = m_{H^+}$. 
In this approach, we find here similarly to $\bar{B}\to X_s\gamma$ a strong correlation between $\tan\beta$ and $m_{H^+}$.
Later in our global fit (Section~\ref{sec:global_fit}), we will discuss further the dependence on these additional parameters. 

\begin{figure}[t]
    \centering
    \includegraphics[width=0.48\textwidth]{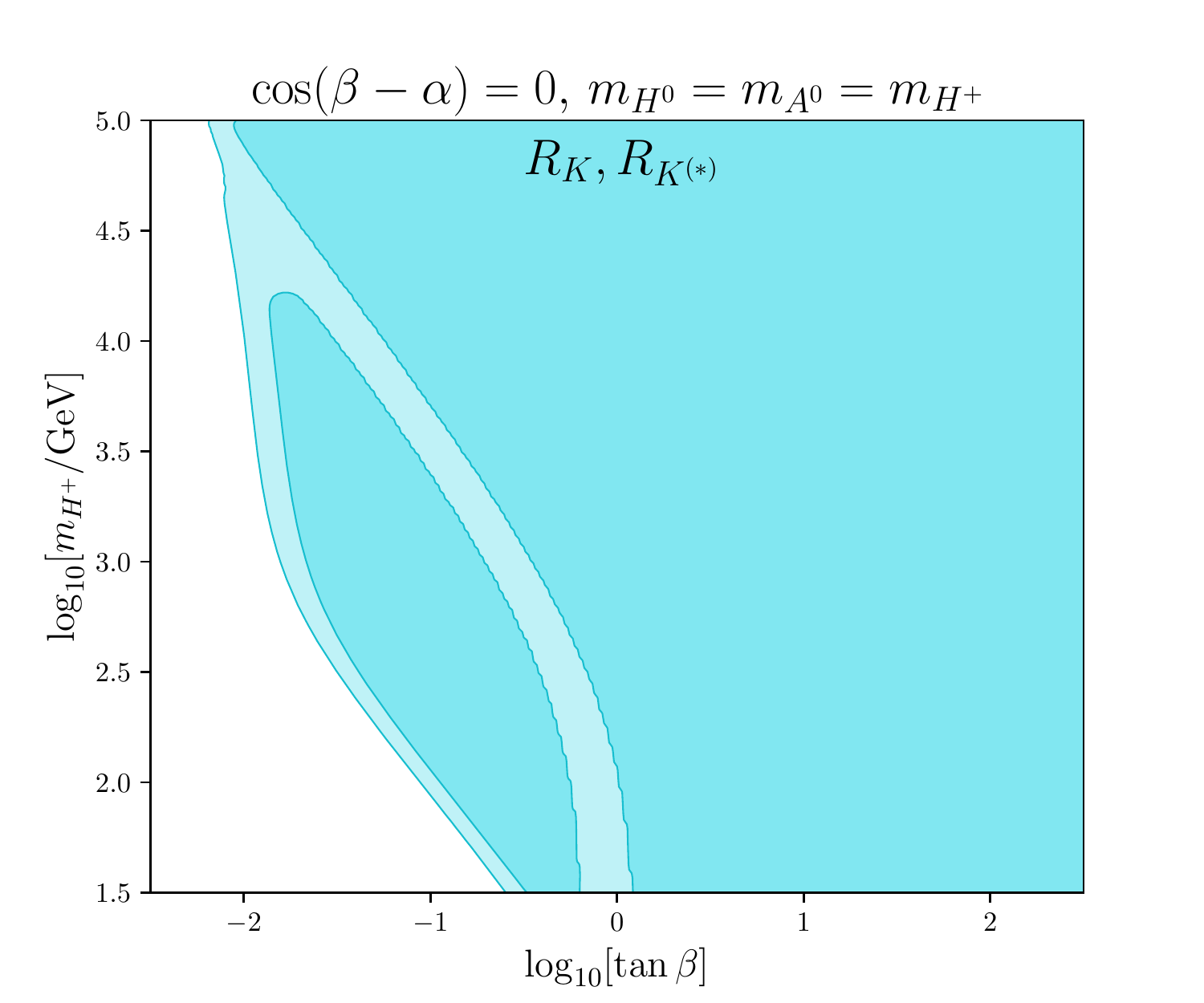}
    \caption{Contour plot of the allowed 2HDM-I parameter space in the ($\tan\beta-m_{H^+}$) plane for the combined fit to the 10 $R_{K^{(*)}}$ bins in Table, taken in the limits of alignment ($\cos(\beta-\alpha)=0$) and degenerate masses ($m_{H^0}=m_{A^0}=m_{H^+}$). The darker contour indicates allowed parameter space at $1\sigma$ confidence, and the lighter at $2\sigma$. }
    \label{fig:rks}
\end{figure}

The LFU ratios $R_K$ and $R_{K^*}$ \cite{Bobeth:2007dw} have become well-studied processes in recent years (see e.g. Ref.~\cite{Bordone:2016gaq,Hurth:2020ehu,Geng:2017svp,Ciuchini:2019usw,Datta:2019zca,Kowalska:2019ley,Hurth:2020rzx,Ciuchini:2020gvn}), 
driven by LHCb measurements \cite{LHCb:2014vgu,LHCb:2017avl} finding discrepancies with the SM, most noticeably the deviation of $3.1\sigma$ in $R_{K^+}$ \cite{LHCb:2021trn}.
Model-independent analyses (for example, Refs.~\cite{Alguero:2019ptt,Alok:2019ufo,Aebischer:2019mlg,Altmannshofer:2017yso}) 
show these processes favour vector-like new physics contributing to ${\cal O}_{9,10}$ for their deviations to be resolved. 
In Table~\ref{tab:Comb_fit_res} we show the best fit point only including the 10 $R_{K^{(*)}}$ bins we consider, where we find that this fit favours the limits $\tan\beta\to\infty$, $\cos(\beta-\alpha)\to0$ and $m_{H^+}\sim m_{H^0}\sim m_{A^0} \sim 50\,$GeV.
This result is purely numerics-driven; we note that it lies outside the bounds of our model given by theoretical considerations.
In the large $\tan\beta$ limit we recover the SM predictions, as the 2HDM-I induced coupling deviations will go to zero. 
With the SM result equating to our best fit point, no point in our parameter space can reduce the tensions in these observables beyond their current status. 
In Fig.~\ref{fig:rks}, we show our combined fit for the 10 $R_{K^{(*)}}$ bins within a more similar parameter region to other fits, where we see that most of our parameter space considered lies within $1\sigma$ of the best fit (or SM) result. 
A discovered 2HDM-I within the physical bounds of the model could only increase the tensions in $R_{K^{(*)}}$ as we find currently them in the SM.

\begin{figure}[!t]
    \centering
    \includegraphics[width=0.48\textwidth]{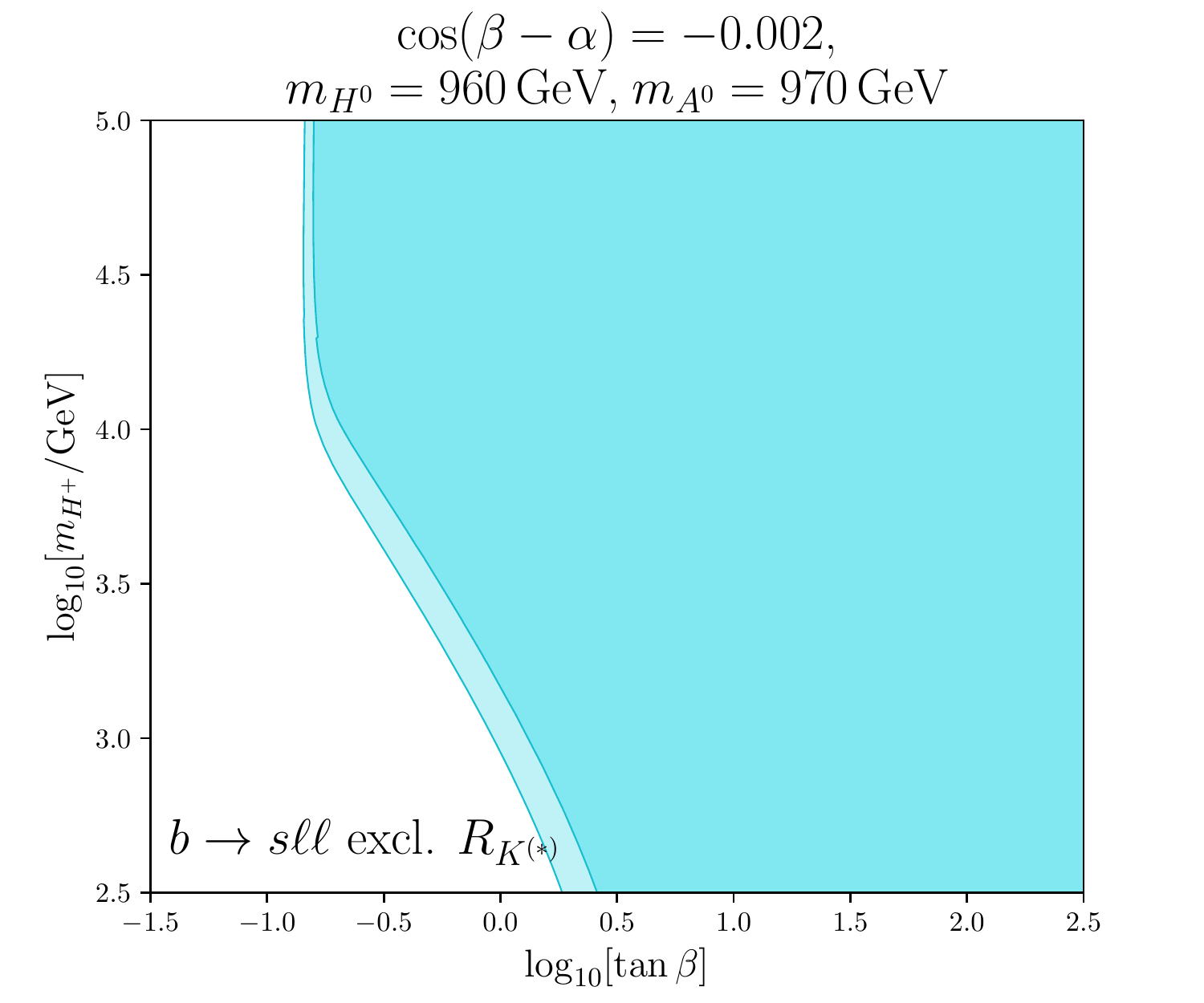}
    \caption{ Contour plot of the allowed 2HDM-I parameter space in the ($\tan\beta-m_{H^+}$) plane for the combined fit to the leptonic $B_{d,s}\to\mu^+\mu^-$  decays and the semi-leptonic $b\to s\ell^+\ell^-$ observables, excluding $R_{K^{(*)}}$ and fixing the additional parameters to their best-fit points. The darker contour indicates allowed parameter space at $1\sigma$ confidence, and the lighter at $2\sigma$.}
    \label{fig:bsll}
\end{figure}

In Fig.~\ref{fig:bsll}, we consider more $b\to s\ell^+\ell^-$ observables (see Table 7 in Ref.~\cite{Atkinson:2021eox}), where some of these also find tension with experiment in the SM. 
For detailed descriptions and analyses focused specifically on this interesting group of observables, see e.g. Refs.~\cite{Khodjamirian:2010vf, Khodjamirian:2012rm, Bharucha:2015bzk, Khodjamirian:2017fxg, Alguero:2019ptt, Hurth:2020ehu, Alok:2019ufo, Ciuchini:2019usw, Hurth:2020rzx, Hurth:2021nsi,Alguero:2021anc,Cornella:2021sby,Altmannshofer:2021qrr,Geng:2021nhg,Ciuchini:2020gvn,MunirBhutta:2020ber,Biswas:2020uaq}.
Similarly to $R_K$ and $R_{K^*}$ above, these processes tend to favour new physics (NP) contributions from ${\cal O}^{(')}_{9,10}$.
In the 2HDM-I, we find that the contributions to these operators (and also to ${\cal O}^{(')}_{S,P}$) can be sufficient to reduce the tension with experiment for some of these observables. 
This effect is most significant for $m_{H^0}\sim m_{A^0}\sim m_{H^+}\sim1000\,$GeV as suggested by the best-fit point in Table~\ref{tab:Comb_fit_res}, where we find an improvement over the SM with a pull of $2.6\sigma$ (excluding $R_{K^{(*)}}$); we then find a large portion of our parameter space within $1\sigma$ confidence of this result. 
Varying the fit parameters around the best fit point, we also find that this group of observables imposes a lower bound on $\tan\beta$ as
\begin{equation}
    \tan\beta \gtrsim 0.14\,(0.16)\quad {\rm at}\;\: 2\sigma\,(1\sigma),
\end{equation}
which is compatible with the lower bound coming from perturbativity in Section~\ref{sec:theory}.

\subsection{Global Fits}
\label{sec:global_fit}
\begin{figure}[t]
    \centering
    \includegraphics[width=0.48\textwidth]{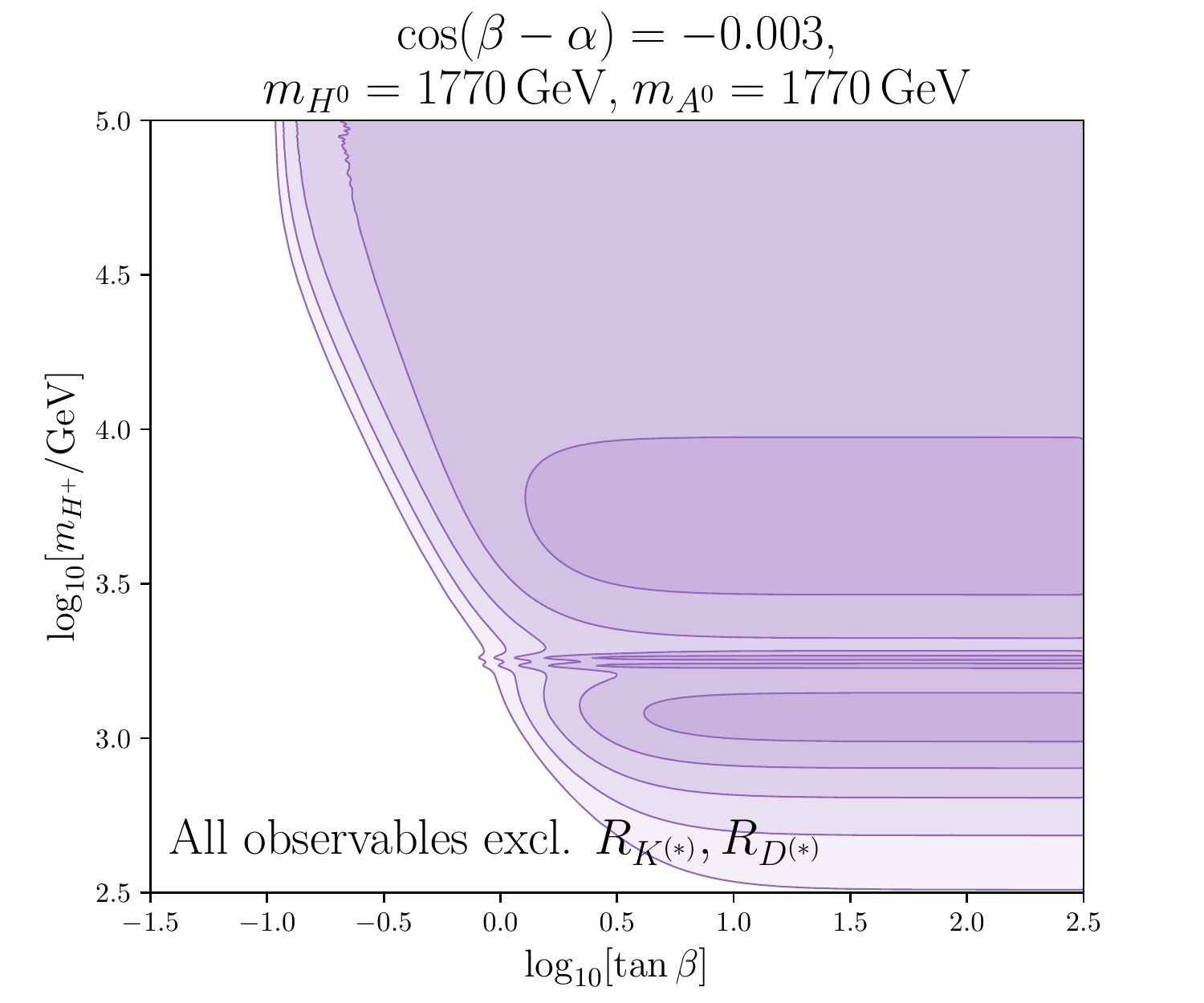}
    \includegraphics[width=0.48\textwidth]{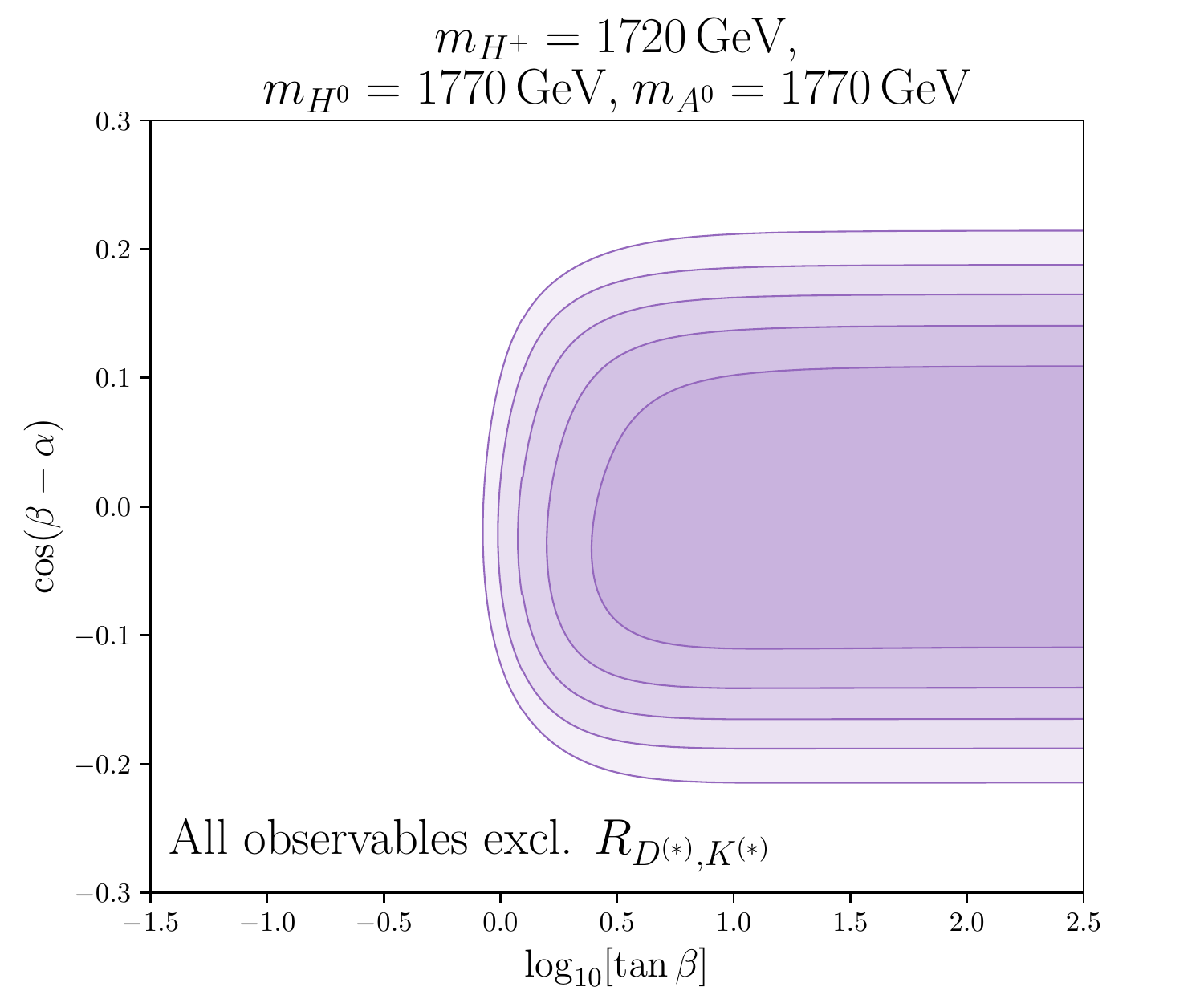}
    \caption{Combined fit of all flavour observables (excluding \LFU), Higgs signal strengths, and EWPOs in the 2HDM-I (in the $\tan\beta-m_{H^+}$ plane), fixing the additional parameters to their best-fit points. 
    Contours are shown representing allowed parameter space at 1,2,3,4,5$\sigma$ confidence from darkest to lightest.}
    \label{fig:global_exrk}
\end{figure}

\begin{figure}[t]
	\centering
	\includegraphics[width=0.48\textwidth]{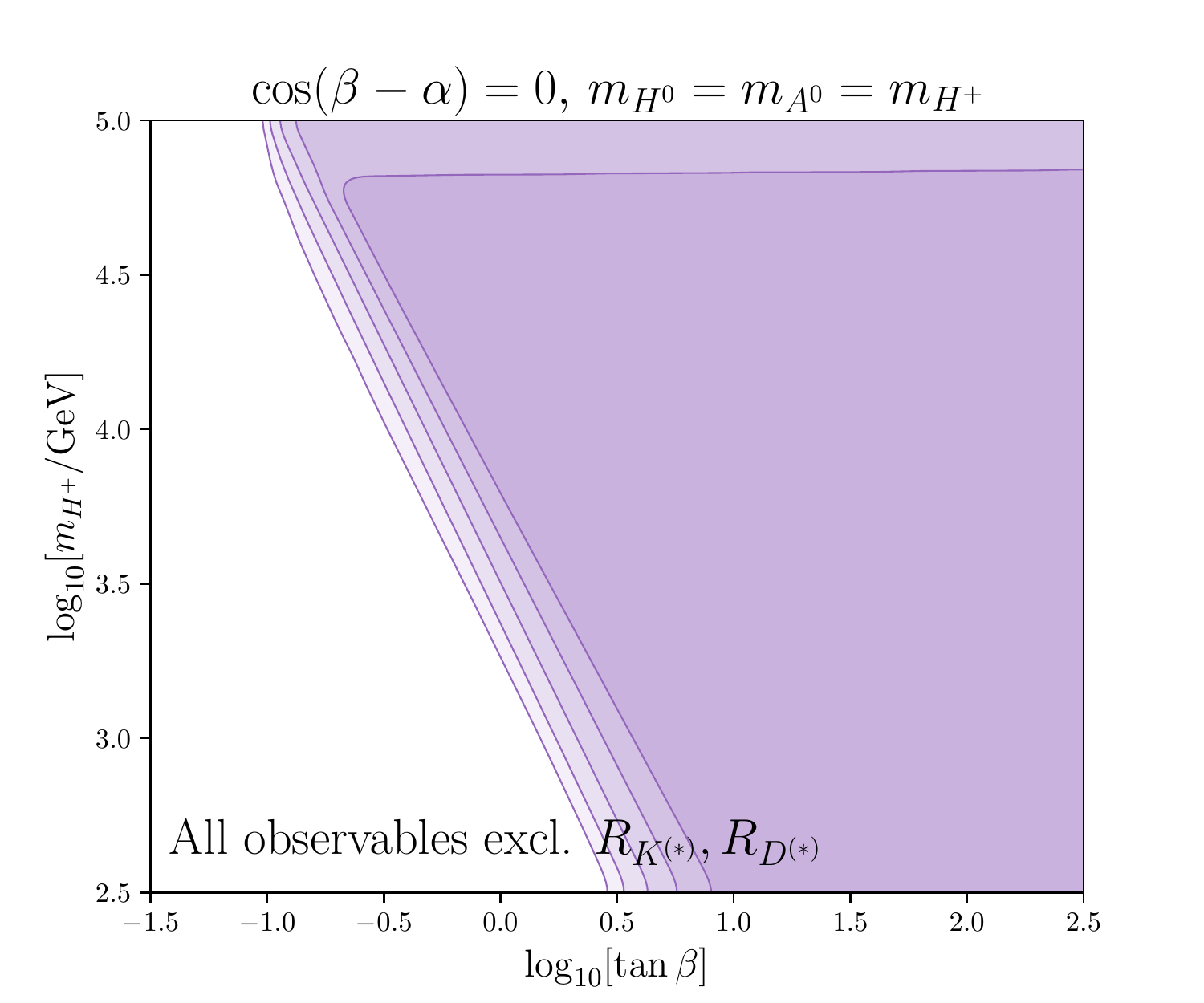}
	\caption{Combined fit of all flavour observables (excluding \LFU), Higgs signal strengths, and EWPOs in the 2HDM-I (in the $\tan\beta-m_{H^+}$ plane), taken in the limits of alignment ($\cos(\beta-\alpha)=0$) and degenerate masses ($m_{H^0}=m_{A^0}=m_{H^+}$). Contours are shown representing allowed parameter space at 1,2,3,4,5$\sigma$ confidence from darkest to lightest.}
	\label{fig:global_exrk_lim}
\end{figure}

We now consider our global fits to the 2HDM-I combining flavour observables, Higgs signal strengths, and the EWPOs $S,T,U$. 
Within the contours of a global fit, the LFU ratios $R_{D^{(*)}}$, $R_{K^{(*)}}$ still have a tension $\sim3.5\sigma$ with experiment that is worsened compared to the SM.\footnote{Some other $b\to s\ell\ell$ observables still find some tension, but these are not as severe and are lessened compared to the SM.}
This motivates two scenarios for the global fits: either including these LFU ratios, or considering a fit excluding them.
In Table~\ref{tab:Comb_fit_res}, we summarise the results of the various fits performed, indicating their best-fit parameter points and the corresponding statistics.
In Fig.~\ref{fig:global_exrk}, we show the fit to all observables (excluding \LFU) where we fix the additional 2HDM parameters to their best fit point as shown in Table~\ref{tab:Comb_fit_res}.
In Fig.~\ref{fig:global_exrk_lim}, we show the fit to all observables (excluding \LFU) in the alignment limit and with degenerate masses. 
We find the shapes of the contours to be much the same for excluding/including \LFU, so we do not show both sets of plots here; the impact of including \LFU \, is clearer in Table~\ref{tab:Comb_fit_res}, where we find distinctly poorer quality of fit (low $p$-value) in this scenario.
Excluding \LFU, we find a pull of $1.9\sigma$ improvement from the SM to 2HDM-I; including \LFU, the improvement is $1.5\sigma$.

From Table~\ref{tab:Comb_fit_res}, we see that $\cos(\beta-\alpha)=0$ (the alignment limit) is closely favoured from the fits, although Fig.~\ref{fig:global_exrk} shows that larger deviations from alignment are allowed than in the 2HDM-II.
We find preference in the global fits for $\tan\beta\sim{\cal O}(100)$ and $m_{H^+}\sim m_{H^0}\sim m_{A^0}\sim 1.7\,$TeV, where a closer mass degeneracy at this scale is enforced from theory constraints. 
Within the parameter region found we scan around the best fit point to find the full constraints on our parameters.
As one can see throughout this section, much of the small $\tan\beta$ parameter space is excluded by most observables individually and globally.
We cannot state an explicit bound on $\tan\beta$ from the global fit however as the lower limit is clearly correlated with the charged Higgs mass.
For the mass of the charged Higgs, within the parameter region we test, we find bounds only at $1\sigma$ confidence:
\begin{equation}
    m_{H^+}\leq 83.4\,{\rm TeV} \quad {\rm at}\; 1\sigma.
\end{equation}
In our global fit of all observables, our constraints on $\cos(\beta-\alpha)$ are much improved from considering the Higgs signal strengths alone, where we find the constraints in the full analysis to follow very closely to the representative plot in Fig.~\ref{fig:global_exrk}:
\begin{equation}
\label{eq:cba_lims}
\begin{aligned}
|\cos(\beta-\alpha)| & \leq 0.11 & \quad  {\rm at} &\; 1\sigma, \\
|\cos(\beta-\alpha)| & \leq 0.14 & \quad  {\rm at} &\; 2\sigma, \\
|\cos(\beta-\alpha)| & \leq 0.17 & \quad  {\rm at} &\; 3\sigma, \\
|\cos(\beta-\alpha)| & \leq 0.19 & \quad  {\rm at} &\; 4\sigma, \\
|\cos(\beta-\alpha)| & \leq 0.21 & \quad  {\rm at} &\; 5\sigma.
\end{aligned}
\end{equation}
\renewcommand{\arraystretch}{1.3}
\begin{table}[t] 
	\centering
	\begin{footnotesize}
		\begin{tabular}{|l|c|c|c|c|}
			\hline
			Scenario &  $\#$ & Best-fit point &
			$\chi_{min}^2$ & $p$-value \\
			& Observables & $\{\tan\beta,m_{H^+},m_{H^0},m_{A^0},\cos(\beta-\alpha)\}$ & & \\
			\hline
			{All incl. \LFU}
			& {$275$}
			& {$\{ 70, 1720\,{\rm GeV}, 1680\,{\rm GeV}, 1670\,{\rm GeV}, -0.001 \}$}
			& {$317.0$}
			& {$\phantom{0}2.6\%$} \\
			{All excl. \LFU}
			& {$263$}
			& {$\{ 80, 1720\,{\rm GeV}, 1770\,{\rm GeV}, 1770\,{\rm GeV}, -0.003 \}$}
			& {$281.1$}
			& {$15.5\%$} \\
			{Flavour incl. \LFU}
			& {$241$}
			& {$\{ 60, 1650\,{\rm GeV}, 1600\,{\rm GeV}, 1600\,{\rm GeV}, -0.002 \}$}
			& {$289.1$}
			& {$\phantom{0}1.0\%$} \\
			{Flavour excl. \LFU}
			& {$229$}
			& {$\{ 30, 1790\,{\rm GeV}, 1750\,{\rm GeV}, 1740\,{\rm GeV}, -0.002 \}$}
			& {$249.4$}
			& {$11.7\%$} \\
			{$b\to s\ell\ell$ incl. $R_{K^{(*)}}$}
			& {$202$}
			& $\{ 40, 1030\,{\rm GeV}, 920\,{\rm GeV}, 1010\,{\rm GeV}, -0.003 \}$
			& {$262.1$}
			& {$\phantom{0}0.1\%$} \\
			{$b\to s\ell\ell$ excl. $R_{K^{(*)}}$}
			& {$192$}
			& $\{ 35, 1020\,{\rm GeV}, 960\,{\rm GeV}, 970\,{\rm GeV}, -0.002 \}$
			& {$234.4$}
			& {$\phantom{0}1.1\%$} \\
			{Only $R_{K^{(*)}}$}
			& {$10$}
			& {$\{ >1000, 50\,{\rm GeV}, 60\,{\rm GeV}, 50\,{\rm GeV}, 0.0001 \}$}
			& $\phantom{0}27.9$
			& $0.004\%$ \\
			{Higgs Signals}
			& {$31$}
			& {$\{ >1000, --, --, --, 0.0 \}$}
			& $\phantom{0}28.2$
			& $50.7\%$ \\
			\hline\hline
			
			\multicolumn{5}{|c|}{Alignment Limit, $\cos(\beta-\alpha)=0$} \\
			\hline
			{All incl. \LFU}
			& {$275$}
			& {$\{ 40, 1760\,{\rm GeV}, 1730\,{\rm GeV}, 1690\,{\rm GeV}\}$}
			& {$318.5$}
			& {$\phantom{0}2.5\%$} \\
			{All excl. \LFU}
			& {$263$}
			& {$\{ 50, 1810\,{\rm GeV}, 1750\,{\rm GeV}, 1760\,{\rm GeV}\}$}
			& {$283.7$}
			& {$13.9\%$} \\
			\hline
			\hline
			\multicolumn{5}{|c|}{$\cos(\beta-\alpha)=0$, $m_{H^+} = m_{H^0} = m_{A^0}$} \\
			\hline
			{All incl. \LFU}
			& {$275$}
			& {$\{ 90, 1630\,{\rm GeV}\}$}
			& {$324.3$}
			& {$\phantom{0}1.8\%$} \\
			{All excl. \LFU}
			& {$263$}
			& {$\{ 80, 1750\,{\rm GeV}\}$}
			& {$286.4$}
			& {$13.9\%$} \\
			\hline
		\end{tabular}
	\end{footnotesize}
	\caption{Best fit points of 2HDM-I parameter fits for various groups of observables using the constraints from theory to inform the physical parameter values.}
	\label{tab:Comb_fit_res}
\end{table}

\subsection{Comment on the Muon Anomalous Magnetic Moment}
\begin{figure}[th]
    \centering
    \includegraphics[width=0.48\textwidth]{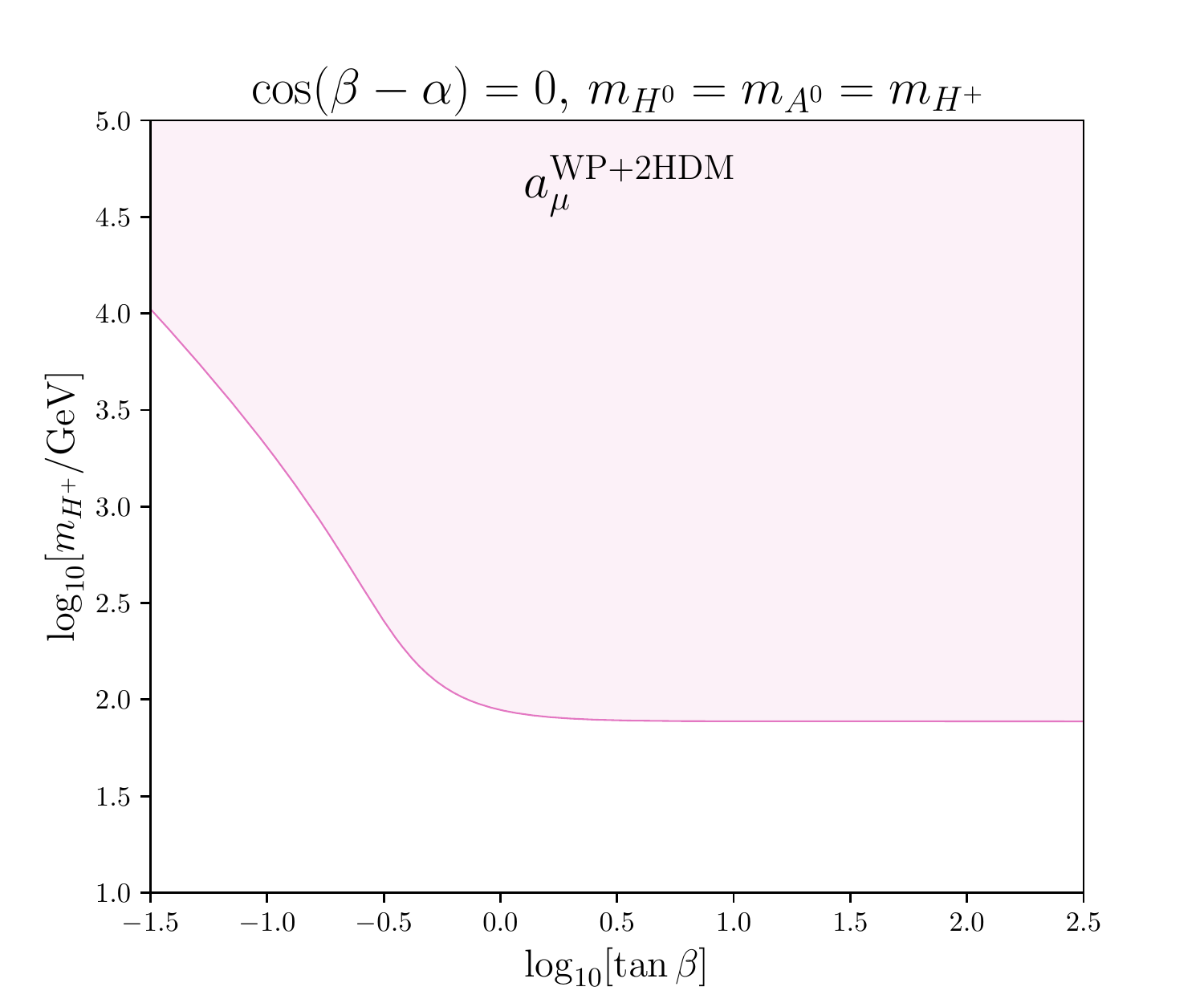}
    \includegraphics[width=0.48\textwidth]{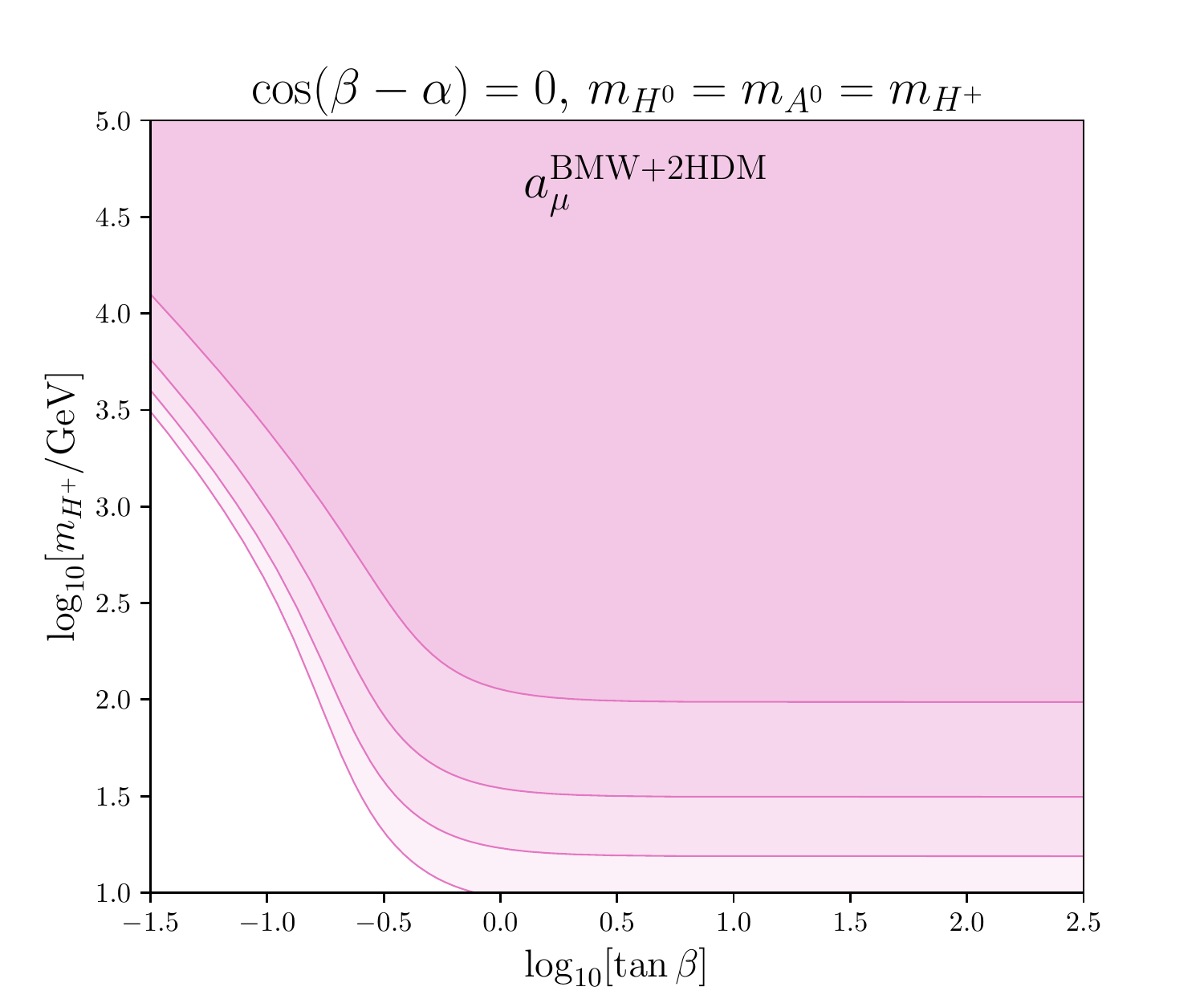}
    \caption{Contour plot of allowed 2HDM-I parameter space in the ($\tan\beta-m_{H^+}$) plane for $a_\mu$, taken in the limits of alignment ($\cos(\beta-\alpha)=0$) and degenerate masses ($m_{H^0}=m_{A^0}=m_{H^+}$). In the left plot, the SM prediction taken from the theory initiative is used; in the right, the SM result from the BMW collaboration. Contours are plotted representing allowed parameter space at 1,2,3,4,5$\sigma$ confidence from darkest to lightest: in the left plot, only the $5\sigma$ contour is visible; in the right, the $2,3,4,5\sigma$ contours.}
    \label{fig:g-2}
\end{figure}

It is worthwhile to comment on the anomalous magnetic moment of the muon, $a_\mu$, which we do not include in our overall fit.
The recent results from Run 1 at Fermilab \cite{Muong-2:2021ojo} confirmed the previous result from BNL \cite{Muong-2:2006rrc}, where the combined experimental value now yields a $4.2\sigma$ deviation from the SM result predicted by the Theory Initiative White Paper (WP)~\cite{Aoyama:2020ynm} (based on~Refs.~\cite{Aoyama:2012wk,Aoyama:2019ryr,Czarnecki:2002nt,Gnendiger:2013pva,Davier:2017zfy,Keshavarzi:2018mgv,Colangelo:2018mtw,Hoferichter:2019mqg,Davier:2019can,Keshavarzi:2019abf,Kurz:2014wya,Melnikov:2003xd,Masjuan:2017tvw,Colangelo:2017fiz,Hoferichter:2018kwz,Gerardin:2019vio,Bijnens:2019ghy,Colangelo:2019uex,Blum:2019ugy,Colangelo:2014qya}).
There is also a competing Lattice QCD prediction for the $a_\mu^{\rm HVP}$ contribution from the BMW Collaboration~\cite{Borsanyi:2020mff} which results in only a $1.6\sigma$ discrepancy from experiment.\footnote{There are several lattice calculations \cite{FermilabLattice:2017wgj,Budapest-Marseille-Wuppertal:2017okr,RBC:2018dos,Giusti:2019xct,Shintani:2019wai,FermilabLattice:2019ugu,Gerardin:2019rua,Aubin:2019usy,Giusti:2019hkz} which were included in the WP, however the precision of these are much lower than either data-driven approach or the BMW calculation.}
Here, we fit $a_\mu$ to the 2HDM-I parameters considering both the WP and the BMW SM scenarios, shown in Fig.~\ref{fig:g-2}.
We take the one- and two-loop contributions of the 2HDM to $a_\mu$ from Ref.~\cite{Ilisie:2015tra} where one can insert the expressions for the 2HDM-I couplings and we then convert the results to {\bf flavio}'s WET-3 basis as described in Section 7.2 of Ref.~\cite{Atkinson:2021eox}.
As before, to present information in terms of two-dimensional fits, we fix the additional 2HDM parameters $\cos(\beta-\alpha)=0$ and $m_{H^+}=m_{H^0}=m_{A^0}$ as motivated by theory. 

We find that in the 2HDM-I, $a_\mu$ strongly favours low $m_{H^+}\sim100\,$MeV, low $\tan\beta\sim0.5$, where the tension between experiment and theory using either SM prediction can be reduced to much less than $1\sigma$. 
This result is however below the physical domain for $m_{H^+}$. 
Within the physical domain, we find that $m_{H^+}\gtrsim100\,$GeV can yield a tension between the WP and experiment of less than $5\sigma$, with a minimum of $4.3\sigma$, and between BMW and experiment less than a $2\sigma$ tension, with a minimum of $1.6\sigma$.

\subsection{2HDM-I and II Flavour Prospects at Future Colliders}
\label{sec:Future_flav}
Finally, we would like to discuss the prospects of the 2HDM flavour sector at future colliders. 
An approximate prediction for B-physics at future colliders such as the HL-LHC or beyond is that the precision of the measurements will roughly double from that of today's \cite{Cerri:2018ypt,FCC:2018byv}. 
With this in mind, we explore several scenarios for flavour observables in the future. 
Denoting the present-day experimental value of some observable as $M \pm E$, where $M$ is the central value and $E$ the uncertainty, for each observable we now set $E \to E/2$ 
and consider 5 scenarios for a shift in $M$ in the range $[M-E,M+E]$, where 
$M$ for all observables shifts in the same way.
We also consider an ``Ideal'' future scenario whereby $M$ for each observable is individually shifted within $[M - E, M + E]$ to minimise its best-fit point $\chi^2$ contribution; this scenario generally brings $M$ to the SM prediction or as close as possible within $[M - E, M + E]$, and can be considered qualitatively similar to an extrapolation assuming the SM result.

We perform this process on $\sim$200 of our 275 observables which are identified as being \\
``experimentally-limited" in their $\chi^2$ contribution: that is, their present $\chi^2$ contribution is dominated by the size of their experimental uncertainty. 
It is assumed in these tests that the theoretical calculations and precision do not change significantly by the time of these future measurements.

In Table~\ref{tab:future_flavour}, we list the comparison of the $\chi^2$ best fits in the SM, 2HDM-I, 2HDM-II for first the present situation and then each future scenario considered. 
We also list the lower bound of $m_{H^+}$ from $\bar B\to X_s\gamma$ in the 2HDM-II for each scenario, as this has been an important bound in the history of the 2HDM and can remain so in the future. 
It is unlikely that any one of the exact scenarios would come to pass from future measurements for such a large group of observables, however comparison between all 6 scenarios sheds some light on the future suitability of the SM and the 2HDMs. 
In the present, we find that both the 2HDM-I and 2HDM-II perform better than the SM by $\sim2\sigma$. 
In all future scenarios, the 2HDM-II still performs better than the SM, with a pull $>1\sigma$ in all but one of these. 
The 2HDM-I's performance over the SM is narrowed significantly however, where the best future scenarios for this have a reduced pull of $1\sigma$. 
In the ``Ideal'' future, the SM, 2HDM-I, and 2HDM-II all significantly improve upon their current positions; the majority of observables in this ``Ideal'' scenario simply move much closer to their SM predictions, where the small contributions from the 2HDMs either are not significant or are sufficient to improve upon the remaining small tensions with the SM. 

Overall, it is found that increased precision of flavour observables at future colliders can lead to much poorer fits in both the SM and the 2HDM. 
The exception to this is when measurements are shifted much closer to the SM predictions and the small 2HDM contributions can be sufficient to resolve much of the remaining differences.
While the future measurements' central values may not follow the scenarios considered here, the limiting factor in the $\chi^2$ values in Table~\ref{tab:future_flavour} is the increased precision of measurements. 
Unless future measurements would indeed conform closely to our ``Ideal'' future scenario, the performances of the SM, 2HDM-I, 2HDM-II would all become very poor.
While assuming no significant gaps in our theory predictions, this suggests some New Physics other than the 2HDM should be explored.
{\renewcommand{\arraystretch}{1.3}
\begin{table}[th]
    \centering
    \begin{tabular}{|c|c|c|c|c|}
        \hline\hline
        Scenario & SM & 2HDM-I & 2HDM-II & Min($m_{H^+}$)$^{b\to s\gamma}_{\rm 2HDM-II},2\sigma$ \\
        \hline
        Present & 292 & 281 (+$1.9\sigma$) & 282 (+$1.8\sigma$) & 790 GeV \\
        \hline
        ``Ideal'' future & 174 & 169 (+$0.8\sigma$) & 166 (+$1.4\sigma$) & 680 GeV \\
        \hline
         $M  \pm E/2$ & 721 & 717 (+$0.7\sigma$) & 711 (+$1.7\sigma$) & 940 GeV \\ 
         \hline
        $(M-E) \pm E/2$ & 811 & 808 (+$0.4\sigma$) & 808 (+$0.4\sigma$) & 1050 GeV \\
        $(M-E/2) \pm E/2$ & 815 & 812 (+$0.4\sigma$) & 804 (+$2.0\sigma$) & 1000 GeV \\
        $(M+E/2) \pm E/2$ & 896 & 891 (+$0.8\sigma$) & 887 (+$1.6\sigma$) & 680 GeV \\
        $(M+E) \pm E/2$ & 899 & 893 (+$1.0\sigma$) & 887 (+$2.1\sigma$) & 770 GeV \\ 
        \hline \hline
        & \multicolumn{4}{c|}{Best-fit point $\{ \tan \beta, m_{H^+}, m_{H^0}, m_{A^0}, 
        \cos (\beta - \alpha) \}$} \\
        \hline
        2HDM-I & \multicolumn{4}{c|}{$\{ 80, 1720\,{\rm GeV}, 1770\,{\rm GeV}, 1770\,{\rm GeV}, -0.003 \}$} \\ 
        \hline
        2HDM-II & \multicolumn{4}{c|}{$\{ 4.3, 2340\,{\rm GeV}, 2380\,{\rm GeV}, 2390\,{\rm GeV}, 0.009 \}$} \\
        \hline\hline
    \end{tabular} 
    \caption{Summary of future prediction scenarios and their $\chi^2$ best fits. For the 2HDMs in brackets are the pulls from the SM where a positive value indicates an improvement and a negative value a worsening. The results are shown for excluding \LFU. Also shown is the variation of the lower bound of the charged Higgs mass in the 2HDM-II from $\bar B\to X_s\gamma$.}
    \label{tab:future_flavour}
\end{table}}

\section{LHC Constraints, Flavour Comparison and Outlook}
\label{sec:lhc}
Having explored flavour constraints and their projections in detail in the previous section, we now turn to constraints from collider measurements of exotics searches. Such direct searches for Higgs bosons have a long history. Indeed, searching for the Higgs boson of the SM was the raison d'être of the LHC, picking up where other colliders such as the LEP left off. In the years since the Higgs discovery, many precise measurements of the 125 GeV scalar have been made (see above), while searches for other, more exotic, Higgs bosons have continued alongside the precision measurements of the observed Higgs (see also~Refs.~\cite{Chen:2013rba,Chen:2013kt}). These precision measurements closely match the phenomenology predicted of the SM Higgs. In this Section we look to combine the wealth of experimental data on direct searches for new Higgs bosons to further constrain the allowed parameter space of the 2HDM, independently from the constraints in the prior sections, before combining these results in Section~\ref{sec:overlaps}. We also extrapolate the current LHC bounds to the expected results at the HL-LHC operating at 3~ab$^{-1}$ in the same channels.

As in the rest of this work, the parameters of primary interest here are the masses of the new Higgs bosons and the angles $\tan\beta$ and $\cos(\beta-\alpha)$. Following the results of Section~\ref{sec:global_fit}, we examine the favoured scenario in which the new Higgs bosons have degenerate masses and the alignment limit holds.

\subsection{Current Collider Bounds}
\label{sec:LHC_0}

\begin{figure}[th]
    \centering
    \includegraphics[width=0.48\textwidth]{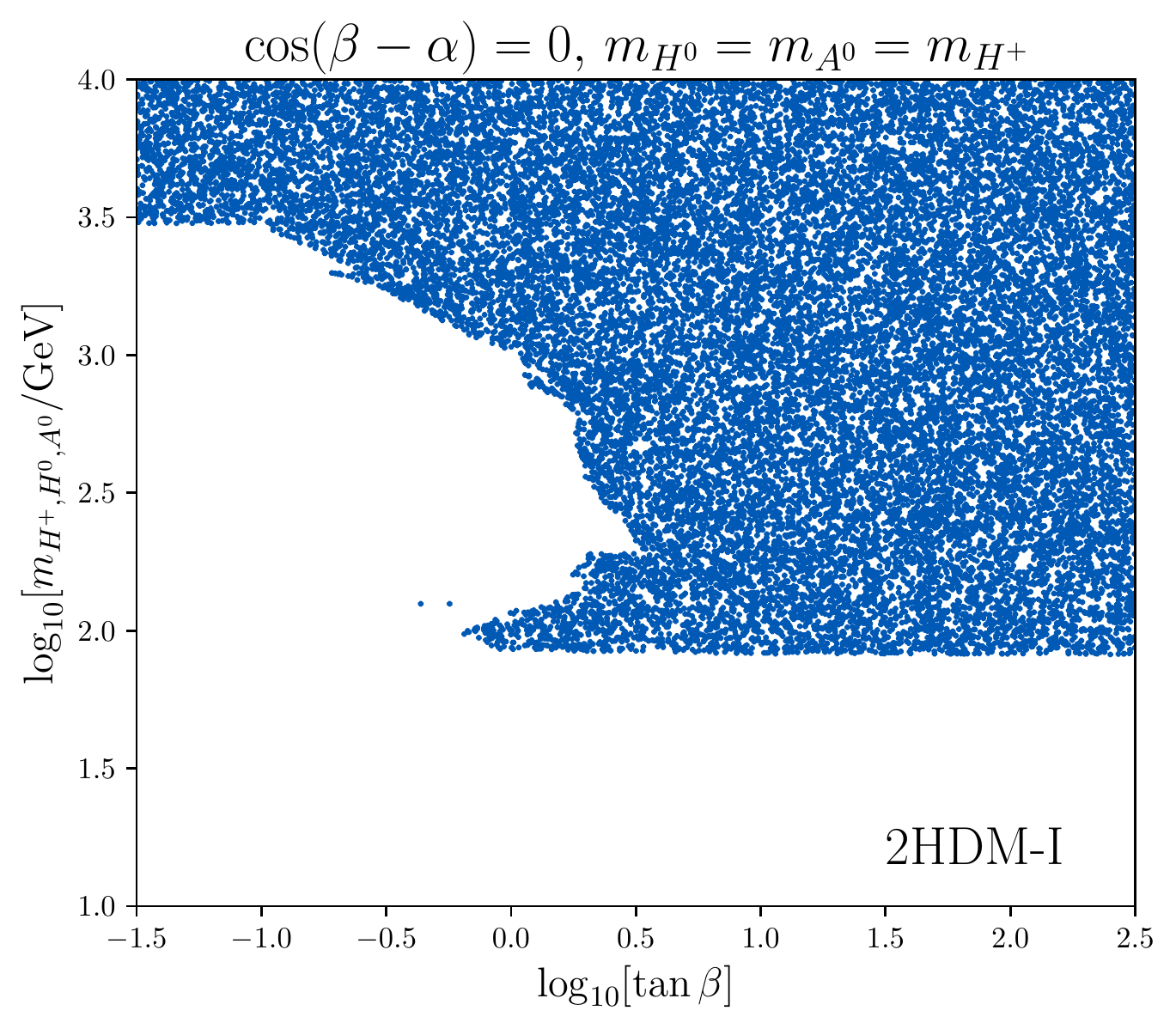}
    \includegraphics[width=0.48\textwidth]{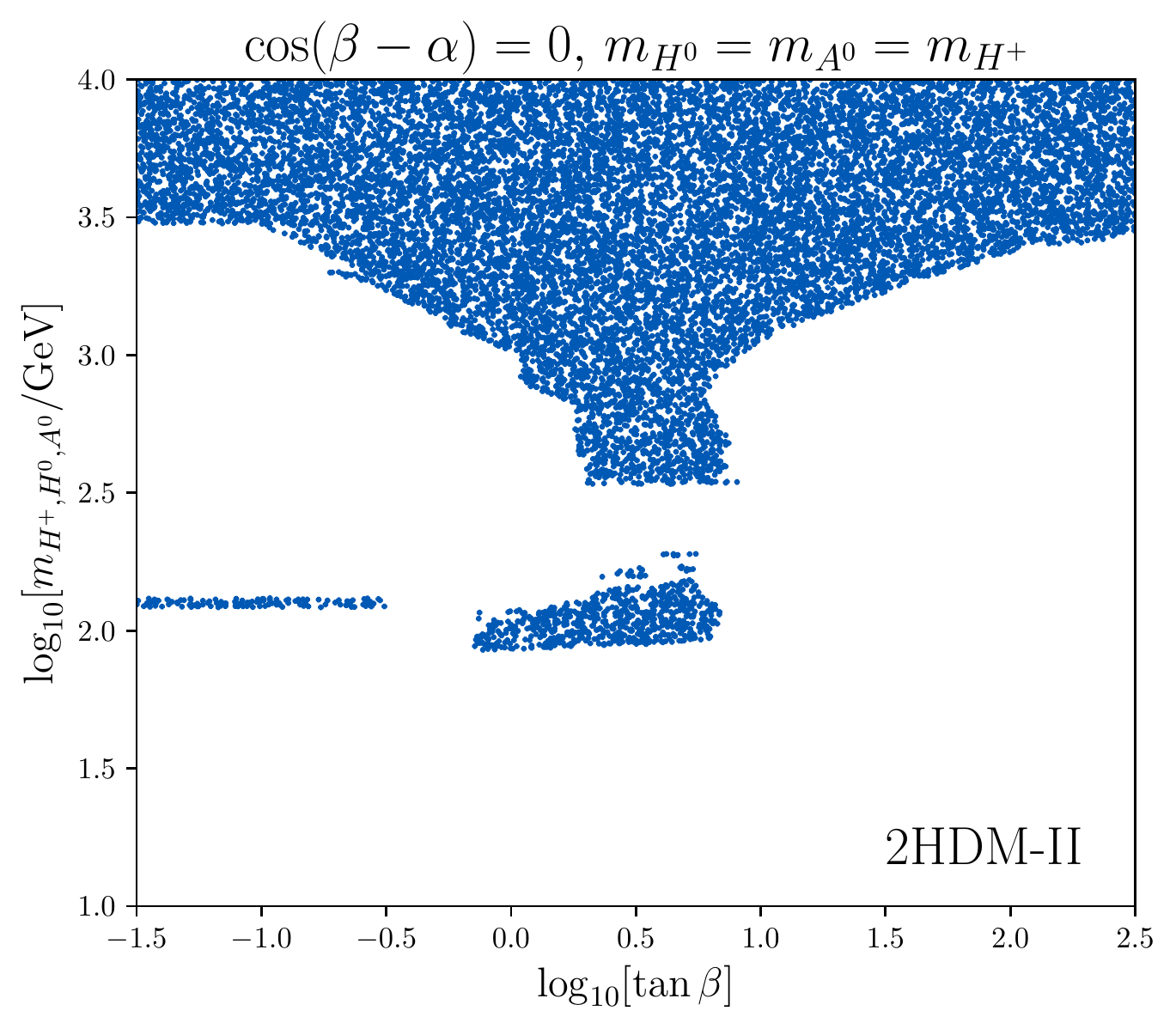}
    \caption{Scan of 50000 randomly generated points compared against LHC data using \texttt{2HDecay} and \texttt{HiggsBounds}, with allowed points shown in blue. On the left, the scan in the 2HDM-I, on the right the 2HDM-II.}
    \label{fig:LHC_0}
\end{figure}

In order to leverage the power of the vast array of search data we use the packages \texttt{HiggsBounds} \cite{Bechtle:2008jh, Bechtle:2011sb, Bechtle:2012lvg, Bechtle:2013wla, Bechtle:2015pma, Bechtle:2020pkv, Bahl:2021yhk} and \texttt{2HDecay} \cite{Krause:2018wmo, Djouadi:1997yw, Djouadi:2018xqq, Krause:2016xku, Denner:2018opp, Hahn:1998yk}. We use \texttt{2HDecay} to calculate the branching ratios and decay widths of each 2HDM Higgs boson for a given point in parameter space, and interface this data with \texttt{HiggsBounds} to check if the search data excludes such a point. For the key channel of $H^+$ production in association with $t\bar{b}$, we use \texttt{MadGraph5\_aMC@NLO} \cite{Alwall:2014hca} to generate cross sections which also form part of the input to \texttt{HiggsBounds}, along with the couplings of the Higgs bosons as given in Table~\ref{tab:xis}. We perform scans using 50000 randomly generated points in the parameter space following this method and show the allowed points for both type I and II 2HDMs in Fig.~\ref{fig:LHC_0}. The historic baseline sensitivity from LEP \cite{ALEPH:2013htx} excludes charged Higgs masses below 72.5~GeV and 80~GeV for the Type I and Type II 2HDM models respectively.

In the 2HDM-I scan on the left of Fig.~\ref{fig:LHC_0}, at low values of $\tan\beta$, leptonic decays of the neutral Higgses (e.g. $H^0 \to \mu^+\mu^-$ in Ref.~\cite{CMS:2019lwf}) exclude lower masses. Sensitivity to $H^+ \to \tau^+\nu_\tau$ peaks at $\sim 85$ GeV \cite{CMS:2019bfg} and then the $H^+ \to t\bar{b}$ \cite{ATLAS:2020jqj} decay provides the exclusion from $\sim 250$ GeV until falling cross sections lead to a loss of sensitivity at $\sim 2.5$~TeV, beyond which points are allowed. This exclusion limit falls with increasing $\tan\beta$ as the $H^+tb$ coupling is proportional to $\cot^2\beta$.\footnote{The relevant Yukawa coupling is $2((m_t\xi^u_{H^+})^2+(m_b\xi^d_{H^+})^2)/v^2$, where $\xi^{u,d}_{H^+}=\xi^{u,d}_{A}$ in Table \ref{tab:xis}.}  At moderate $\tan\beta$ the low mass exclusion limit shifts to $H^0/A^0 \to 4b$ \cite{ALEPH:2006tnd}. The kink at $\sim 100$ GeV in the moderate $\tan\beta$ region is a result of $\mathcal{B}(H^+\to\tau^+\nu_\tau)$ falling to zero as $\mathcal{B}(H^+\to t\bar{b})$ rises to unity in this mass region, with $\mathcal{B}(H^+\to t\bar{b}) \approx 1$ from $m_{H^\pm} \gtrsim m_t$. Sensitivity is lost from $H^0 \to \ell\ell$ as $\kappa^{H^0}_\ell \propto \cot\beta$, see Table~\ref{tab:xis}. The flat cut off at $80$ GeV results from $H^+ \to qq/\tau^+\nu_\tau$ \cite{ALEPH:2013htx}.

For the case of the 2HDM-II at low $\tan\beta$, $H^0 \to \gamma\gamma$ and $H^+ \to t\bar{b}$ compete to give the most stringent exclusion bounds \cite{ATLAS:2018xad, ATLAS:2014jdv, ATLAS:2020jqj}, with $H^+ \to t\bar{b}$ being more sensitive at higher masses before the production cross sections fall at high masses. At moderate values of $\tan\beta$,  $H^+ \to \tau^+\nu_\tau$ excludes masses up to $\sim$~90 GeV, before the branching ratio for this channel falls, which leads to the allowed region at $\tan\beta$ of order 1 and masses of $\sim$ 100 GeV. This region is then ended by the $H^0 \to \tau^+\tau^-$ search in Ref.~\cite{ATLAS:2020zms}. This search also gives the exclusion up to high masses in the large $\tan\beta$ region, exceeding the limits from $H^+ \to t\bar{b}$. 

We are able to find lower mass bounds on the new Higgs bosons of 82~GeV and 86~GeV in the Type I and II 2HDM respectively, which are less stringent bounds than those found from the combined fit to flavour observables in Section~\ref{sec:global_fit}. These bounds improve at low $\tan\beta$ in both models, and also at high $\tan\beta$ in the 2HDM-II, owing to the dependence of the couplings of these models to $\tan\beta$. This is in line with the patterns we see in the flavour sector, where these couplings are also crucial.

\subsection{Collider Outlook}
\label{sec:LHC_3000}

\begin{figure}[th]
    \centering
    \includegraphics[width=0.48\textwidth]{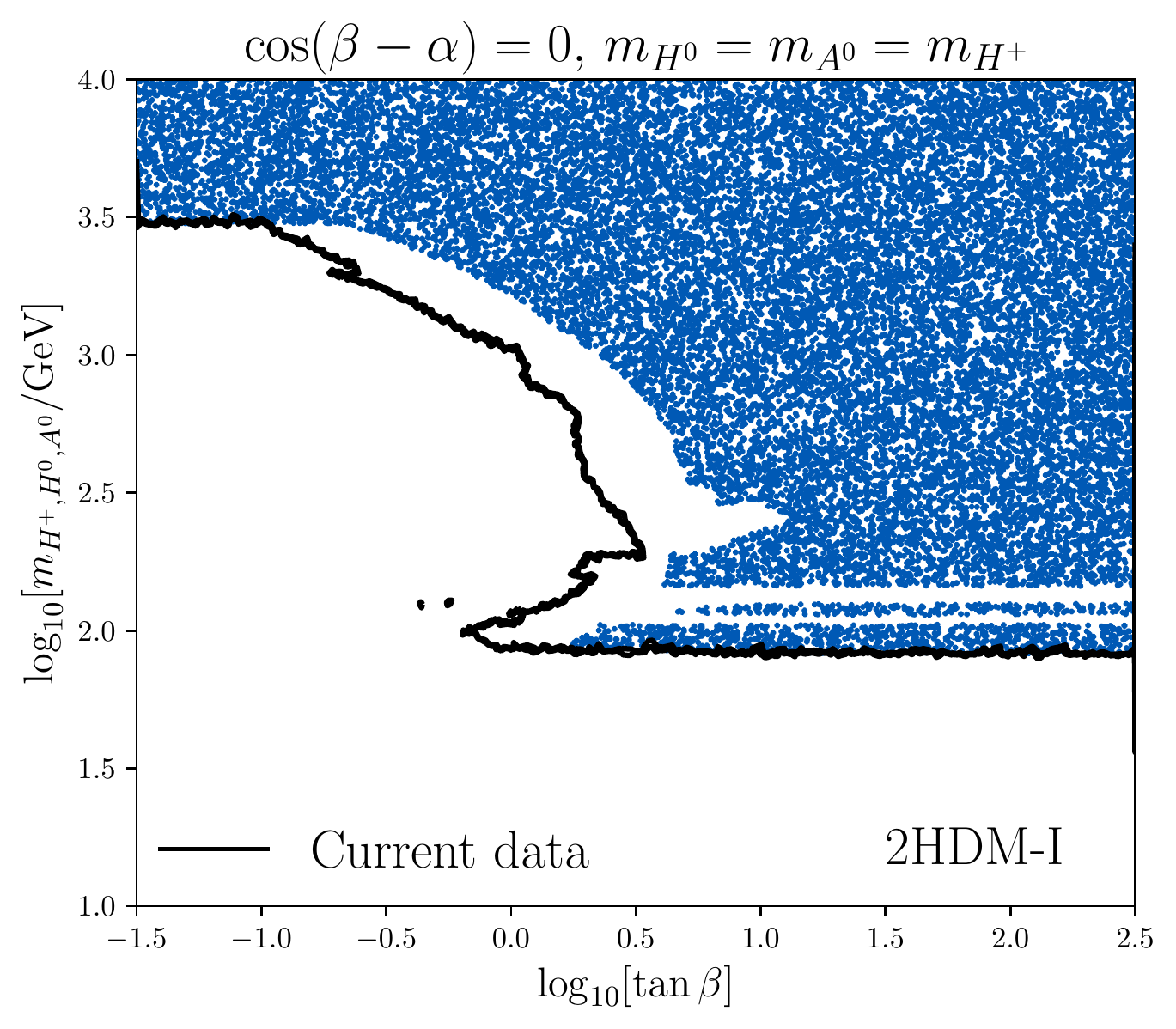}
    \includegraphics[width=0.48\textwidth]{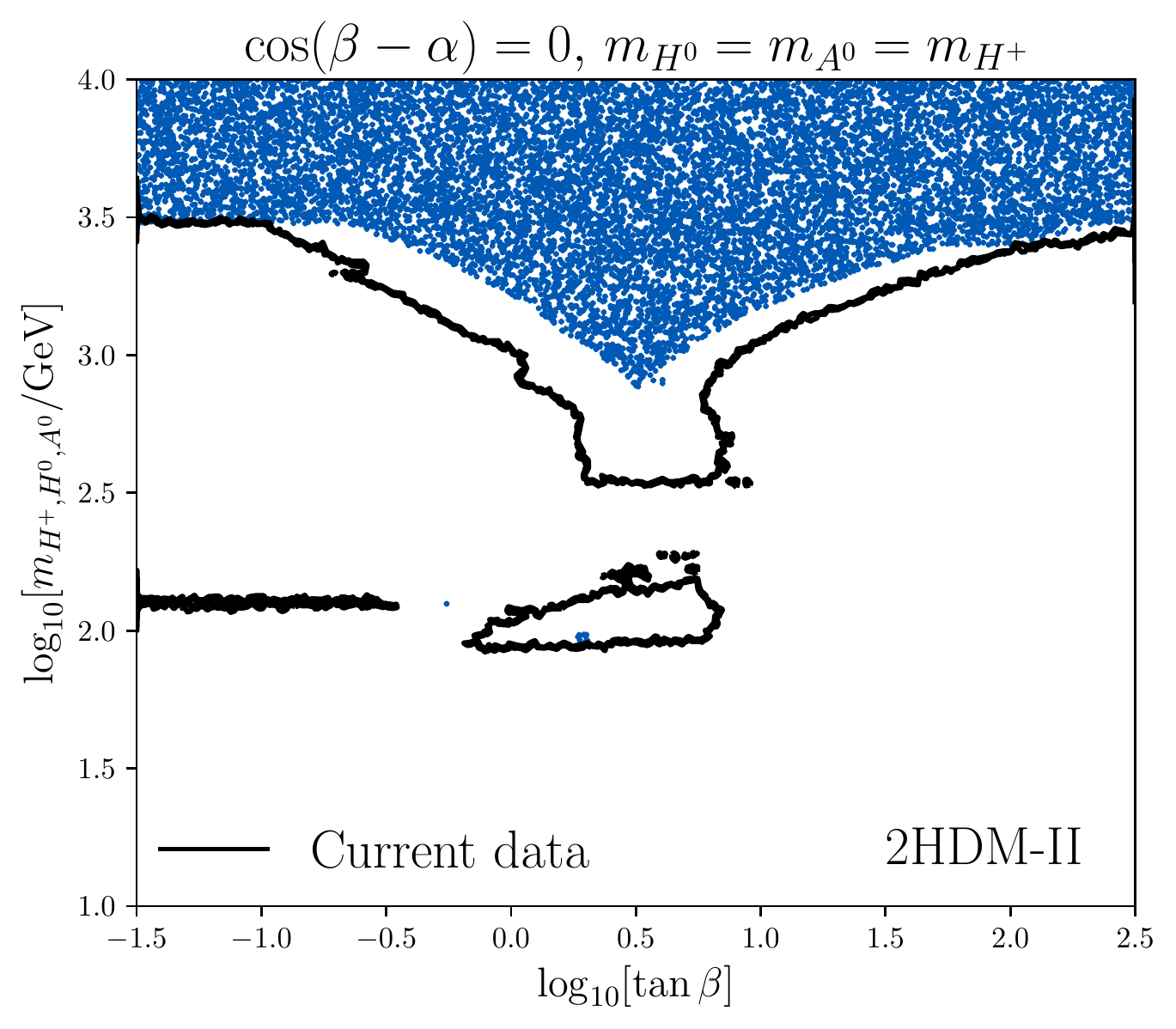}
    \caption{Scan of 50000 randomly generated points compared against LHC data, extrapolated to an integrated luminosity of 3~ab$^{-1}$ using {\tt{2HDecay}} and \texttt{HiggsBounds}, with allowed points shown in blue and the current data exclusion contour in black. On the left, the scan in the 2HDM-I, on the right the 2HDM-II.}
    \label{fig:LHC_3000}
\end{figure}

Looking to the future, the LHC is due to be upgraded considerably, increasing the integrated luminosity to $\mathcal{L}_{\text{HL-LHC}} = 3$~ab$^{-1}$. This will have a sizeable impact, making it significantly harder for new particles to remain hidden. Here, we extrapolate the LHC data currently in \texttt{HiggsBounds} to this new luminosity by scaling the limits by a factor $\sqrt{(\mathcal{L}_0/\mathcal{L}_{\text{HL-LHC}})}$ for a search with a reference luminosity $\mathcal{L}_0$. There are important caveats to this extrapolation. One such caveat is that a number of searches are designed to precisely measure the SM Higgs behaviour. In these cases we match the bounds to what would be expected of the SM Higgs, as here we have a 125 GeV scalar in the exact alignment limit, which exactly matches the phenomenology of the SM Higgs. We make use of the SM Higgs predictions of~Ref.~\cite{LHCHiggsCrossSectionWorkingGroup:2016ypw} for this. Additionally, a number of searches included in \texttt{HiggsBounds} are from LHC run 1, during which the LHC operated at a centre of mass energy of $\sqrt{s} = 7-8$~TeV, whilst the HL-LHC will operate at $\sqrt{s} = 13-14$~TeV. We therefore need to reflect increased cross sections for these searches together with luminosity improvements (see also~Ref.~\cite{Basler:2018dac}). We again make use of the results from Ref.~\cite{LHCHiggsCrossSectionWorkingGroup:2016ypw} to perform this extrapolation for SM Higgs searches. For BSM Higgs searches we use {\tt MadGraph5\_aMC@NLO}~\cite{Alwall:2014hca} to calculate the increase in the production cross sections at the higher centre of mass energy as a function of the BSM Higgs mass and scale the search data limits accordingly. We again perform a scan consisting of 50000 random points, the results of which are shown in Fig.~\ref{fig:LHC_3000}, which includes the outline of the scan in Fig.~\ref{fig:LHC_0} for ease of comparison.

The permitted parameter space for the 2HDM is reduced at the future collider, based on extrapolating existing searches. In the 2HDM-I case this effect is only apparent at low $\tan\beta$, as the sensitivity to new Higgs bosons is minimal at high $\tan\beta$ because the relevant couplings are proportional to $\cot^2\beta$. The lower limit in this region is from a LEP search \cite{ALEPH:2013htx}, which we do not extrapolate here, hence the same cut off is present here as in Fig.~\ref{fig:LHC_0}. There is no such issue in the 2HDM-II, and we find that the bounds improve significantly at the HL-LHC, bringing the lower mass bound into competition with those found from flavour constraints in Section~\ref{sec:Future_flav}, save for a small region with masses of $\approx 95$ GeV and $\tan\beta \approx 2$.

\subsection{Comparison with Flavour}
\label{sec:overlaps}

\begin{figure}[th]
    \centering
    \includegraphics[width=0.48\textwidth]{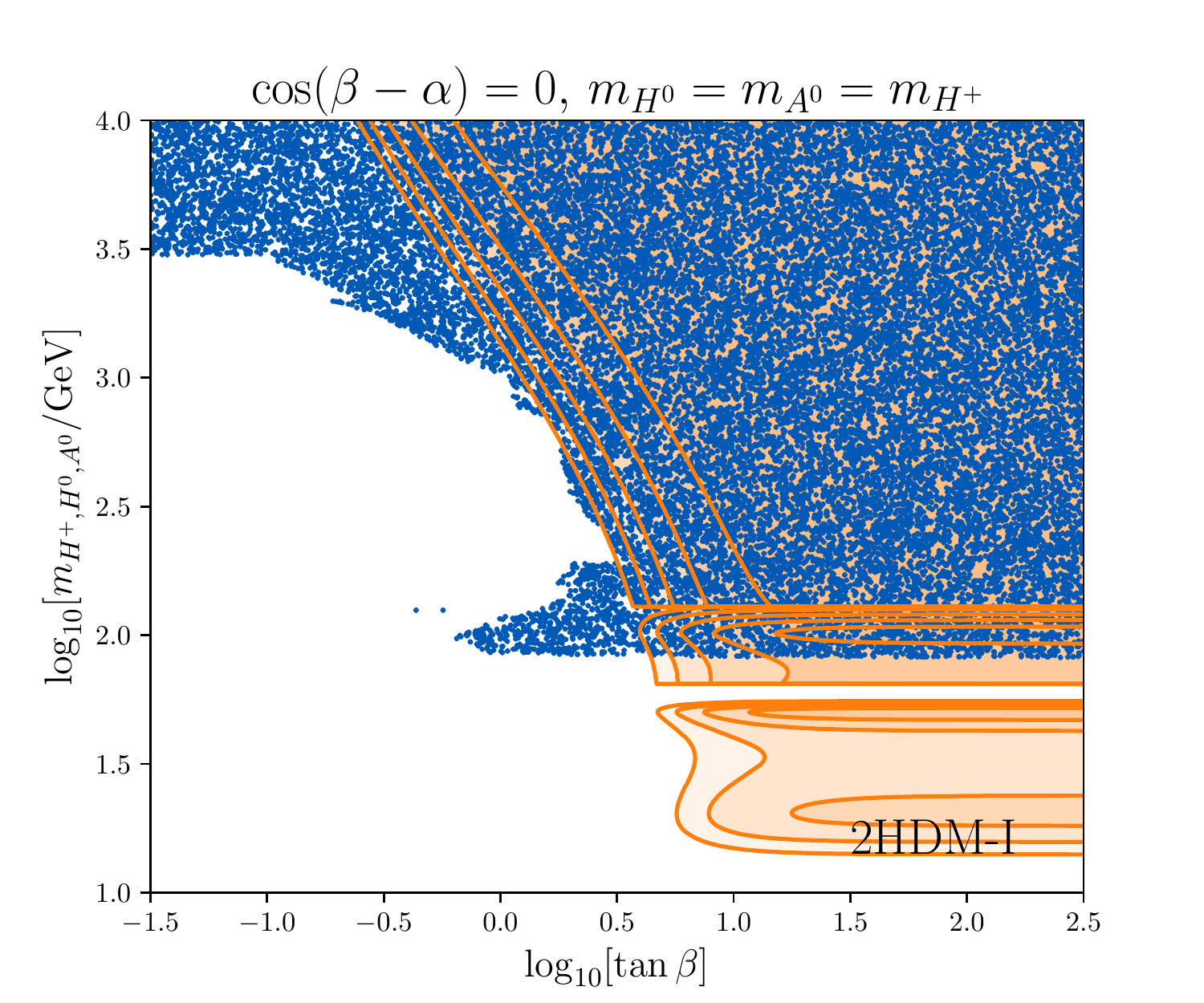}
    \includegraphics[width=0.48\textwidth]{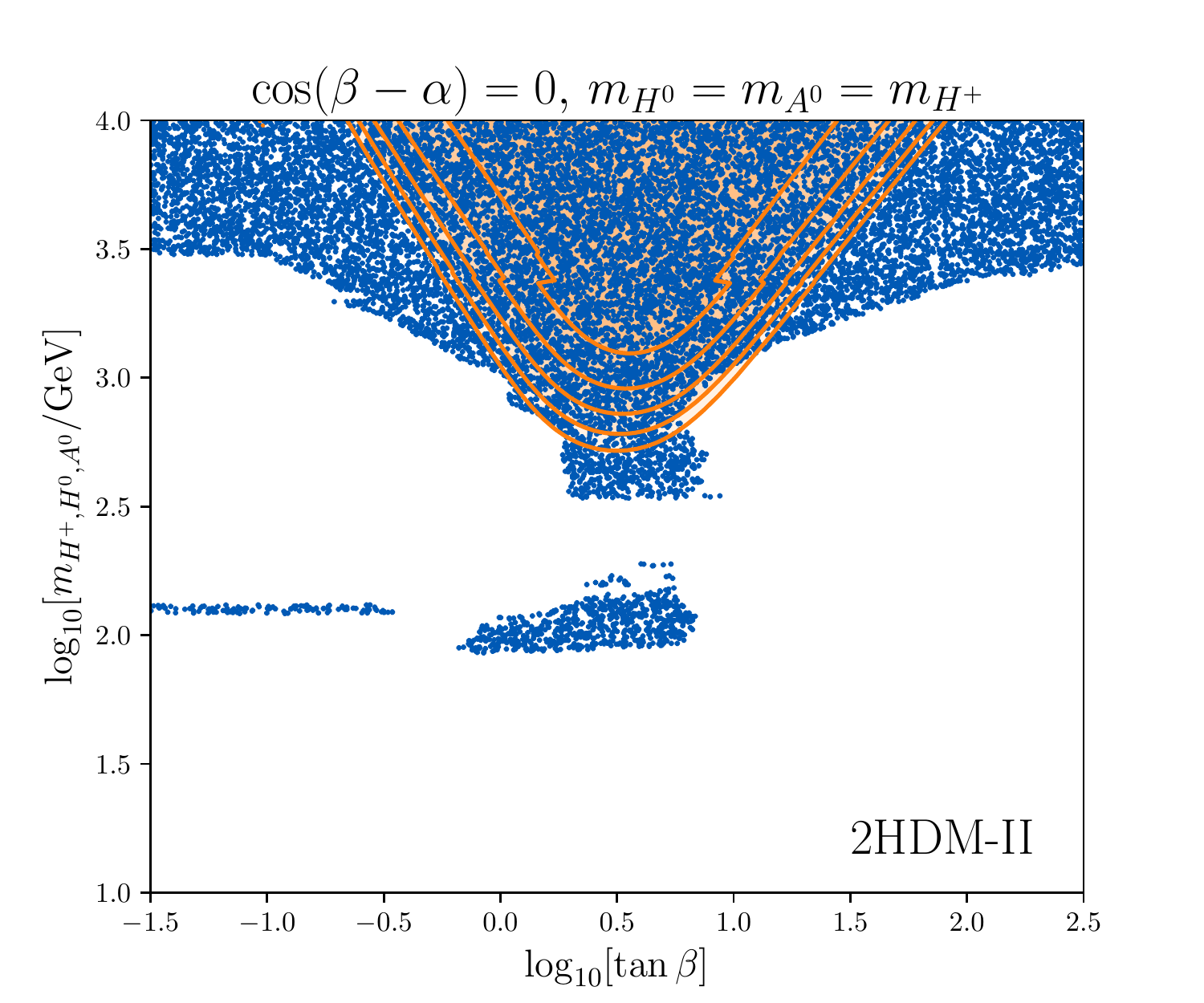}
    \caption{Combined results from the global flavour fit and collider searches, with allowed points from collider searches in blue and contours from the flavour sector at 1,2,3,4,5$\sigma$ confidence, from darkest to lightest, with 2HDM-I results on the left and 2HDM-II on the right.}
    \label{fig:overlaps}
\end{figure}

In order to compare the bounds found above from collider searches to those we find from the flavour sector, we overlap results from both sections. No statistical combination between the two is attempted here as we do not use \texttt{HiggsBounds} to give a combined exclusion from all search data but only check if a point is excluded by any one individual search. We again present the results from the scenario in which all new Higgs bosons have degenerate masses and the alignment limit is exactly realised, meaning the combination presented in Fig.~\ref{fig:overlaps} takes the results from Fig.~\ref{fig:LHC_0} for collider results, and from Fig.~\ref{fig:global_exrk_lim} and the corresponding global fit in Ref.~\cite{Atkinson:2021eox} for flavour results, where we simply extend these contours to lower charged Higgs masses to be more compatible with the collider results.

We find that there is some degree of complementarity between the two sectors for the types of 2HDM that are examined here. In the 2HDM-I the LEP searches \cite{ALEPH:2013htx} that set a lower mass bound on the new Higgses outperform the exclusion from flavour observables, which lack sensitivity in the high $\tan\beta$ region. Conversely, in both cases, the flavour constraints following from precise measurements are more successful constraining $\tan\beta$ at at high masses, reflecting the loss of sensitivity in direct collider searches once the new particles are sufficiently heavy. 
Looking at the contours in Fig.~\ref{fig:LHC_3000}, we can see that the extrapolation to the HL-LHC phase, which improves the sensitivity to new physics, enhances the complementary nature of the results in Fig.~\ref{fig:overlaps}. In the case of the 2HDM-II the lower mass bound may exceed that determined from the flavour sector, depending on how future measurements in that sector line up with the scenarios outlined in Section~\ref{sec:Future_flav}.
\section{A Note on Cosmological Context}
\label{sec:cosmo}
%
As motivated by previous investigations \cite{Dorsch:2014qja,Dorsch:2016tab,Su:2020pjw,Dorsch:2016nrg,Wang:2021ayg,Basler:2016obg,Dorsch:2017nza,Atkinson:2021eox,Basler:2021kgq}, we consider the possibility of generating a strong first order electroweak phase transition (SFOEWPT) through the 2HDM-I in order to fulfil the criterion for EW baryogenesis. Scanning large parameter regions, these find possibilities for a SFOEWPT to be achieved in the 2HDM-I, however this is when considering lower masses ($\lesssim 1\,$TeV) than we find are favoured from our fits.

To evaluate the EWPT in the 2HDM-I at chosen benchmark points, we use the {\bf BSMPT} package \cite{Basler:2018cwe,Basler:2020nrq}, where we refer to the package's documentation\footnote{\href{https://phbasler.github.io/BSMPT/}{https://phbasler.github.io/BSMPT/}} for further information on the package, and we make use of it as discussed in Section 7.3 of Ref.~\cite{Atkinson:2021eox}.
In Table~\ref{tab:bsmpt} we present the results for the strength of the EWPT at our selected parameter points.
We first choose a benchmark of $M=50\,$TeV as a test of extreme masses, and then consider points motivated by our global fits in Table~\ref{tab:Comb_fit_res} as well as some variation in $\tan\beta$. 
As the high masses of our best fits are suggested to be above the limit to achieve a SFOEWPT, we also consider a range of points with masses below $1\,$TeV, including the best fit point for the $b\to s\ell\ell$ observables.
The alignment limit, $\cos(\beta-\alpha)=0$, is taken for all benchmark points.
\begin{table}[th] 
\centering
    \begin{tabular}{|c|ccc|cccc||c|c|c|}
        \hline\hline
        \multirow{2}{*}{$\tan\beta$} & \multicolumn{3}{c|}{Mass Basis (GeV)} & \multicolumn{3}{c}{Lambda Basis} & $m_{12}^2$ & $\omega_c$ & $T_c$ & \multirow{2}{*}{$\xi_c$} \\
        & $m_{H^+}$ & $m_{H^0}$ & $m_{A^0}$ & $\lambda_3$ & $\lambda_4$ & $\lambda_5$ & (GeV$^2$) & (GeV) & (GeV) & \\
        \hline\hline 
        $11.4$ & $50000$ & $50000$ & $50000$ 
            & $0.26$ & $0$ & $0$ & $3.1\times10^{7}$ 
            & $0.58$ & $164$ & $0.004$ \\
            \hline
        $80$ & $1750$ & $1750$ & $1750$ 
            & $0.26$ & $0$ & $0$ & $3.8\times10^{4}$ 
            & $23$ & $162$ & $0.14$  \\
        $50$ & $1810$ & $1750$ & $1760$ 
            & $7.3$ & $-6.5$ & $-0.6$ & $6.1\times10^{4}$ 
            & $26$ & $174$ & $0.15$  \\ 
        $10$ & $1810$ & $1750$ & $1760$ 
            & $7.3$ & $-6.5$ & $-0.6$ & $3.0\times10^{5}$ 
            & $26$ & $174$ & $0.15$ \\
        $150$ & $1810$ & $1750$ & $1760$ 
            & $7.3$ & $-6.5$ & $-0.6$ & $2.0\times10^{4}$ 
            & $26$ & $174$ & $0.15$ \\ 
            \hline
        $35$ & $1020$ & $960$ & $970$ 
            & $4.2$ & $-3.6$ & $-0.3$ & $2.6\times10^{4}$ 
            & $24$ & $169$ & $0.14$  \\
        $80$ & $860$ & $710$ & $860$ 
            & $8.0$ & $-3.9$ & $-3.9$ & $6.3\times10^{3}$ 
            & $142$ & $174$ & $0.82$ \\
        $80$ & $860$ & $690$ & $860$ 
            & $9.0$ & $-4.3$ & $-4.3$ & $6.0\times10^{3}$ 
            & $177$ & $174$ & $1.02$ \\
        $80$ & $680$ & $470$ & $680$ 
            & $8.2$ & $-4.0$ & $-4.0$ & $2.8\times10^{3}$ 
            & $211$ & $147$ & $1.43$ \\
        $80$ & $570$ & $320$ & $570$ 
            & $7.6$ & $-3.7$ & $-3.7$ & $1.3\times10^{3}$ 
            & $226$ & $125$ & $1.81$ \\
        $80$ & $490$ & $250$ & $490$ 
            & $6.1$ & $-2.9$ & $-2.9$ & $7.8\times10^{2}$ 
            & $207$ & $126$ & $1.65$ \\
        $80$ & $490$ & $490$ & $490$ 
            & $0.26$ & $0$ & $0$ & $3.0\times10^{3}$ 
            & $23$ & $161$ & $0.14$ \\
        \hline\hline
    \end{tabular}
    \caption{
    Table of results for the EWPT in the 2HDM-I. $\xi_c=\omega_c/T_c$ is the parameterisation of the strength of the EWPT, with $\omega_c$ the high-temperature VEV and $T_c$ the critical temperature. A SFOEWPT is indicated by $\xi_c>1$. In the limits chosen in Ref.~\cite{Atkinson:2021eox} (and similarly discussed in Ref.~\cite{BhupalDev:2014bir}), $\lambda_1=\lambda_2=m_{h^0}^2/v^2=0.26$ for all benchmark points, and so are not explicitly shown in the table. Additionally, $\lambda_{3,4,5}$ are not independent, but $\lambda_3+\lambda_4+\lambda_5=m_{h^0}^2/v^2=0.26$.}
    \label{tab:bsmpt} 
\end{table}

The results from these benchmark points are similar to those of the 2HDM-II. 
We find that the high mass scales around or above our best fit point (also enforcing close mass degeneracy) limit the strength of the phase transition.
Sufficiently below our best fit points, where theoretical considerations now allow sampling with mass differences similar to those favoured in Ref.~\cite{Su:2020pjw}, we now find scenarios which support a SFOEWPT.
In contrast to the 2HDM-II, these points with $\xi_c\geq1$ are each within $1\sigma$ confidence of our best fit and are allowed by direct search data.
In these scenarios, the global fit performs only slightly worse than the best fit points described in Table~\ref{tab:Comb_fit_res} with improvement over the SM at $\sim1.5\sigma$.

\section{Conclusions}
\label{sec:conclusion}
The search for new physics beyond the Standard Model is key to improving our understanding of physics at the smallest distances. Searches at the LHC for new interactions and states have so far not revealed any significant deviation from the SM expectation, while the precision of flavour and electroweak precision measurements corner the SM from different directions with some potential hints 
for LFU violation.

In this work we have performed a detailed investigation and comparison of the flavour and LHC exotics measurement programme, alongside its future extrapolation. We find that both flavour and current exotics searches can be significantly improved with increasing luminosity, thus allowing the experimental collaborations to cover a significant proportion of parameter space that is currently unconstrained. 

Extending the results of Ref.~\cite{Atkinson:2021eox} for 2HDM-II constraints from indirect searches using electroweak precision, Higgs signal strengths, and flavour observables to the type I 2HDM, we find that the 2HDM-I statistically outperforms both the 2HDM-II and the SM in fits to the data. 
The best fit point for the 2HDM-I lies around
\begin{equation}
\begin{aligned}
    m_{H^+} \approx m_{H^0} &\approx m_{A^0} \approx 1.7 \, \mbox{TeV} \, , \\
    \tan \beta \approx 70, \quad & \, \cos(\beta-\alpha) \approx -0.001  \, .
\end{aligned}
\end{equation}
The parameters of the 2HDM-I are found however to be much less constrained than in the 2HDM-II. 
An upper bound on the charged Higgs mass is only found at $1\sigma$ of $83\,$TeV, where for masses at or above $\sim1\,$TeV, the additional Higgses are expectedly to be closely degenerate.
For a lower charged Higgs mass, $\tan\beta$ is strongly constrained from below to be ${\cal O}(1)$ or higher, however this constraint is lessened with increasing charged Higgs mass. 
The alignment limit, $\cos(\beta-\alpha)=0$, is favoured by our fits, however larger deviations than in the 2HDM-II are allowed, up to $|\cos(\beta-\alpha)|=0.4$ at $5\sigma$.

Expecting an increase in precision of flavour measurements from future colliders, we consider our flavour fits in several future measurements' projections, where these predictive fits encourage consideration of other models instead of (or in addition to) a 2HDM of type I/II to improve upon the SM.

Furthermore, we consider the implications of these flavour fits on the electroweak phase transition of the early universe. 
It is found that the high masses encouraged by our fits do not support a strong EWPT for baryogenesis, however due to the 2HDM-I being less constrained than the 2HDM-II, sufficently low masses are possible within $1\sigma$ of our best fit.

Using data from direct searches for new fundamental particles at colliders we have been able to exclude large regions of the parameter space for the 2HDM of types~I~and~II, highlighting possible areas to search for possible new physics. Looking forward, we have extrapolated a significant amount of data to match the expected performance of the HL-LHC and find that this further improves the constraints on new physics signals, to a level that can be competitive with indirect results from the flavour sector.

While some complementarity between flavour, Higgs physics and direct exotics searches remains, the finite energy coverage of exotics searches typically means a loss of direct LHC sensitivity for large masses approaching the decoupling limit.

\subsection*{Acknowledgements}
O.A. is funded by a STFC studentship under grant ST/V506692/1. 
The work of M.B. is supported by Deutsche Forschungsgemeinschaft (DFG, German Research Foundation) through TRR 257 “Particle Physics Phenomenology after the Higgs Discovery”.
C.E. is supported by the STFC under grant ST/T000945/1, the Leverhulme Trust under grant RPG-2021-031, and the Institute for Particle Physics Phenomenology Associateship Scheme.
Computations carried out for this work made use of the OMNI cluster of the University of Siegen.

\bibliographystyle{JHEP}
\bibliography{references}

\end{document}